\documentclass[fleqn,usenatbib]{rasti}

\usepackage{newtxtext,newtxmath}

\usepackage[T1]{fontenc}

\DeclareRobustCommand{\VAN}[3]{#2}
\let\VANthebibliography\thebibliography
\def\thebibliography{\DeclareRobustCommand{\VAN}[3]{##3}\VANthebibliography}

\usepackage{graphicx}	
\usepackage{amsmath}	
\usepackage{xcolor}
\usepackage{amssymb}
\usepackage[version=4]{mhchem}
\usepackage{chemformula}

\title[Probing Exoplanetary Chemistry with Ariel]{Probing Exoplanetary Chemistry with Ariel: Scientific Priorities and Observational Strategies}

\author[O. Venot {\em et al.\/}]{
\newauthor Olivia Venot,$^{1,2}$
Yamila Miguel,$^{3,4}$
Robin Baeyens,$^{5}$
Stefano Bellotti,$^{6,7}$
Giuseppe Cassone,$^{8}$
\newauthor Quentin Changeat,$^{9}$
Ryan Cloutier,$^{10}$
Athena Coustenis,$^{11}$
Dwaipayan Dubey,$^{12,13}$
Billy Edwards,$^{3}$
\newauthor Kaustubh Hakim,$^{14,15}$
Eric H\'ebrard,$^{16}$\thanks{E-mail: e.hebrard@exeter.ac.uk}
Christiane Helling,$^{17,18}$
Helgi Rafn Hrodmarsson,$^{2}$
Leoni Janssen,$^{4}$
\newauthor Adam Yassin Jaziri,$^{19}$
Gaia Lacedelli,$^{20}$
Panayotis Lavvas,$^{21}$
Jorge Lillo-Box,$^{22}$
Amy Louca,$^{3}$
Adrien Masson,$^{22}$
\newauthor Zita Martins,$^{23}$
Karan Molaverdikhani,$^{12,13,24}$
Benjam\'in Montesinos,$^{22}$
Harrison Nicholls,$^{25}$
Enric Palle,$^{20}$
\newauthor Paul Rimmer,$^{26}$
Donna Rodgers-Lee,$^{27}$
Jonathan Tennyson,$^{28}$
Shang-Min Tsai,$^{29}$
Rom\'eo Veillet,$^{16}$
\newauthor Sergey N. Yurchenko,$^{28}$
and Maria Zamyatina$^{16}$
\\
$^{1}$ Univ. Grenoble Alpes, CNRS, IPAG, 38000 Grenoble, France\\
$^{2}$ Universit\'e Paris Cit\'e and Univ Paris Est Creteil, CNRS, LISA, F-75013 Paris, France\\
$^{3}$ SRON, Netherlands Institute for Space Research, Niels Bohrweg 4, NL-2333 CA, Leiden, The Netherlands\\
$^{4}$ Leiden Observatory, Leiden University, Niels Bohrweg 2, 2333 CA Leiden, The Netherlands\\
$^{5}$ Anton Pannekoek Institute for Astronomy, University of Amsterdam, Science Park 904, 1098 XH, Amsterdam, The Netherlands\\
$^{6}$ Leiden Observatory, Leiden University, PO Box 9513, NL-2300 RA Leiden, The Netherlands\\
$^{7}$ IRAP, Universit\'e de Toulouse, CNRS/UMR 5277, UPS-OMP, 14 Avenue E. Belin, Toulouse F-31400, France\\
$^{8}$ Institute for Chemical-Physical Processes, National Research Council of Italy (IPCF-CNR), 98158 Messina, Italy\\
$^{9}$ Kapteyn Institute, University of Groningen, 9747 AD, Groningen, NL, The Netherlands\\
$^{10}$ Department of Physics \& Astronomy, McMaster University, Hamilton, ON, Canada\\
$^{11}$ LIRA, Paris Observatory, PSL University, CNRS, Paris University, Meudon, 92195, France\\
$^{12}$ Universit\"ats-Sternwarte, Ludwig-Maximilians-Universit\"at M\"unchen, Scheinerstra{\ss}e~1, D-81679 M\"unchen, Germany\\
$^{13}$ Exzellenzcluster Origins, Boltzmannstra{\ss}e 2, 85748 Garching, Germany\\
$^{14}$ Royal Observatory of Belgium, Ringlaan 3, 1180 Brussels, Belgium\\
$^{15}$ Institute of Astronomy, KU Leuven, Celestijnenlaan 200D, 3001 Leuven, Belgium\\
$^{16}$ Department of Physics and Astronomy, Faculty of Environment, Science and Economy, University of Exeter, Exeter EX4 4QL, UK\\
$^{17}$ Space Research Institute, Austrian Academy of Sciences, Schmiedlstra{\ss}e 6, A-8042 Graz, Austria\\
$^{18}$ Graz University of Technology, Graz, Austria\\
$^{19}$ LATMOS/IPSL, UVSQ Universit\'e Paris-Saclay, Sorbonne Universit\'e, CNRS, Guyancourt, France\\
$^{20}$ Instituto de Astrofisica de Canarias (IAC), 38205 La Laguna, Tenerife, Spain\\
$^{21}$ Groupe de Spectrom\'etrie Mol\'eculaire et Atmosph\'erique - (GSMA), Universit\'e Reims Champagne-Ardenne, Reims, 51687, France\\
$^{22}$ Centro de Astrobiolog\'ia (CAB), CSIC-INTA, ESAC campus, Camino Bajo del Castillo s/n, 28692, Villanueva de la Ca\~nada (Madrid), Spain\\
$^{23}$ Centro de Qu\'imica Estrutural, Instituto de Ci\^encias Moleculares e Departamento de Engenharia Qu\'imica, Instituto Superior T\'ecnico, Universidade de Lisboa, 1049-001 Lisboa, Portugal\\
$^{24}$ Max-Planck-Institut f{\"u}r extraterrestrische Physik, Gie{\ss}enbachstr.~1, D-85748 Garching, Germany\\
$^{25}$ Institute of Astronomy, University of Cambridge, Cambridge CB3 0HA, United Kingdom\\
$^{26}$ Cavendish Laboratory, University of Cambridge, Cambridge CB3 0HE, UK\\
$^{27}$ Astronomy \& Astrophysics Section, School of Cosmic Physics, Dublin Institute for Advanced Studies, 31 Fitzwilliam Place, Dublin D02 XF86, Ireland\\
$^{28}$ Department of Physics and Astronomy, University College London, Gower Street, London WC1E 6BT, United Kingdom\\
$^{29}$ Institute of Astronomy \& Astrophysics, Academia Sinica, Taipei 10617, Taiwan
}

\date{Accepted XXX. Received YYY; in original form ZZZ}

\pubyear{\the\year{}}

\begin{document}
\label{firstpage}
\pagerange{\pageref{firstpage}--\pageref{lastpage}}
\maketitle
\clearpage

\begin{abstract}
Over the past two decades, increasingly precise observations have revealed that exoplanet atmospheres are chemically diverse and often far from equilibrium, with processes such as vertical mixing, photochemistry, and atmospheric circulation producing significant departures from thermochemical expectations. As large surveys across a wide range of planets begin to uncover population-level chemical trends, a coherent interpretation of these patterns remains elusive, particularly for gas giants and Neptune-like planets where disequilibrium processes likely dominate. The ESA Ariel mission will provide the first homogeneous, statistically significant atmospheric dataset for nearly a thousand exoplanets, offering an unprecedented opportunity to investigate the origins of chemical diversity across planetary populations. This white paper highlights central scientific questions for understanding atmospheric chemistry and recognized as priorities for the community. These questions span the mechanisms driving disequilibrium chemistry, the role of sulfur- and phosphorus-bearing species, the influence of stellar activity, the formation of haze precursors, and the chemical evolution of atmospheres around stars of different ages and types. Because these topics connect chemistry, physics, and planetary evolution, they form the core focus of the Ariel Chemistry Working Group, which synthesizes current knowledge and identifies the diagnostics best addressed by Ariel's spectral capabilities. For each theme, we outline observational strategies and representative targets already included in the Mission Candidate Sample, illustrating Ariel's ability to address these questions. By linking large-scale observations to predictive atmospheric models, this work supports refinement of Ariel's target selection and enables population-level studies that will transform our understanding of planetary atmospheres.

\end{abstract}




\section{Introduction} 
Over the past two decades, our ability to characterize exoplanet atmospheres has grown dramatically, from the first detections of atoms \citep{Charbonneau2002} to water vapor and more complex molecular species using the space telescopes Hubble \citep{Wakeford2013}, Spitzer \citep{Giovanna2007}, and now JWST \citep{Ahrer2023, Alderson2023, Feinstein2023, Rustamkulov2023,Coulombe2023}. These observations have revealed that exoplanet atmospheres are chemically diverse and often far from equilibrium. For gas giants in particular, deviations from thermochemical equilibrium have emerged as a key diagnostic of atmospheric physics, with vertical mixing, photochemistry, and global circulation all contributing to significant compositional disequilibria \citep{Drummond2020, Venot2020_wasp43b, Tsai2023b}. Over the past decade, large population surveys combining space-based and ground-based data have begun to reveal statistical trends in atmospheric composition and diversity across exoplanets \citep[e.g.][]{Sing2016, Tsiaras2018, Min2020, Changeat2022, Estrela2022, Edwards2023, Deming2023, Saba2025}. These studies have laid the groundwork for population-level atmospheric science, demonstrating the value of consistent and homogeneous analyses across diverse targets. Yet, despite extensive modeling and targeted observations, we still lack a coherent picture of how these processes shape chemical outcomes across the giant planet population.

The ESA Ariel mission has the potential to transform our understanding of planetary atmospheres beyond the Solar System. Scheduled for launch in 2030, Ariel will observe the atmospheres of around 1000 exoplanets across a broad range of sizes, temperatures, and host star types \citep{Tinetti2018, Tinetti2021}. By focusing on a statistically significant and diverse sample, Ariel will go beyond individual detections to search for population-level trends in atmospheric chemistry, composition, and structure.

Ariel's payload consists of a 1.1 m $\times$ 0.73 m off-axis Cassegrain telescope feeding two integrated instruments: the Fine Guidance System (FGS) and the ARIEL InfraRed Spectrometer (AIRS). The FGS includes three photometric channels (0.50-1.10 $\mu$m) and a low-resolution near-infrared spectrometer (NIRSpec, 1.10-1.95 $\mu$m, R $\geq$ 15), while AIRS covers two mid-infrared channels (1.95-3.9 $\mu$m, R $\geq$ 100; 3.9-7.8 $\mu$m, R $\geq$ 30). The payload is passively cooled to $\sim$55 K, with active cooling ($<$ 42 K) only for AIRS detectors, and is optimized for transit spectroscopy from an L2 orbit.

The Ariel mission adopts a four-tier observational strategy to address its scientific objectives. Tier 1 (Reconnaissance Survey) involves low spectral resolution observations of approximately 1000 planets in the visible and infrared (VIS \& IR) with a signal-to-noise ratio (SNR) of $\sim$7, focusing on broad questions such as cloud coverage, atmospheric retention, and bulk properties. Tier 2 (Deep Survey) consists of higher spectral resolution observations of a subsample of planets in the VIS-IR range, aiming to determine main atmospheric components, trace gas abundances, thermal structure, and cloud characterization. Tier 3 (Benchmark Planets) includes high SNR observations of 1-2 events per target, re-observed over time, to study weather, temporal variability, and detailed planetary chemistry and dynamics. Finally, Tier 4 (Phase-curves \& Bespoke Observations) targets specific planets of interest, focusing on phase-curves and bespoke observations to investigate spatial variability and achieve a very detailed understanding of planetary chemistry and dynamics. This tiered approach ensures a comprehensive exploration of exoplanet atmospheres, from broad population studies to in-depth characterizations.

The Chemistry Working Group (WG) within the Ariel mission scientific consortium aims to understand the chemical processes that occur in exoplanet atmospheres using Ariel's infrared spectral capabilities to uncover the mechanisms responsible for chemical diversity across these worlds \citep{Venot2018}. Key topics addressed in the Ariel Chemistry WG include disequilibrium chemistry including different chemical networks, reactions and pathways, the effect of stellar activity on photochemistry, and the elemental ratios that trace planetary evolution and formation histories. 

This white paper outlines the science cases that motivate the Chemistry WG of Ariel. It highlights the need to resolve current ambiguities in chemical trends, develop predictive atmospheric models, and exploit Ariel's unique ability to conduct large-sample atmospheric surveys. For each science case, we present an observation strategy, with a list of target exoplanets. This work is therefore intended to support the selection of targets to be observed.
These efforts will connect observations to theory and help place our Solar System in a broader galactic context.


\section{Giant and Neptune-like planets}
Giant and Neptune-like exoplanets are the primary targets of the Ariel mission, given the mission's instrumental capabilities, particularly its sensitivity to warm and large atmospheres. As a result, most of the science cases discussed in this document are focused on these types of planets. In this Section, we highlight the main scientific objectives related to giant and Neptune-like exoplanets, outlining how Ariel will contribute to our understanding of their formation, composition, and atmospheric dynamics.


\subsection{What is the relationship between the chemical composition of exoplanetary atmospheres and their physical parameters?}\label{sec:population-case1}

\subsubsection{Context}
In the last two decades, the field of exoplanetary science has made remarkable progress, particularly in the detection and atmospheric characterization of exoplanets. Among the key questions in this domain is how the chemical makeup of planetary atmospheres, especially metallicity and C/O ratios, relates to the planets' physical properties and formation histories. Recent population-level analyses have started to explore such relationships empirically, identifying tentative correlations between atmospheric properties and bulk planet parameters \citep[e.g.][]{Sing2016, Tsiaras2018, Min2020, Changeat2022, Estrela2022, Edwards2023, Deming2023, Saba2025}. However, these surveys remain limited in scope and often biased toward hot Jupiters with favorable observing conditions, underscoring the need for a dedicated, statistically representative mission such as Ariel.

As retrieval techniques and atmospheric models become increasingly sophisticated, particularly with the inclusion of non-equilibrium chemistry and photochemical processes \citep[e.g.][Sect.~\ref{sec:disequilibrium}]{Tsai2017,alrefaie2024, bardet2025, changeat2025}, there is a growing expectation that we may soon predict the composition of exoplanetary atmospheres from physical properties of the planets and their host stars. However, such predictive capacity critically depends on establishing robust empirical trends, such as how metallicity, C/O and other ratios scale with planetary mass, radius, or equilibrium temperature.

Currently, our understanding of these relationships remains incomplete. For instance, while some studies suggest a mass-metallicity trend \citep{Welbanks2019}, the statistical evidence is not yet strong due to small sample sizes and large retrieval uncertainties \citep{Sun2024}. Similarly, although C/O ratios are considered potential tracers of formation pathways, especially in relation to snow lines \citep{Oberg2011}, observed atmospheric C/O values span a wide range and often lack a clear correlation with system architecture or stellar composition \citep{Sharma2024}. Some of this scatter is likely attributable to retrieval and model uncertainties: degeneracies between molecular abundances, clouds/hazes, and thermal structure, as well as differences in retrieval frameworks and opacity data, can broaden posterior distributions and contribute to the wide range of inferred C/O values \citep[e.g.][]{Barstow2022,Changeat2022,alrefaie2024,Gao2023,Kempton2023}. Ariel's broad spectral coverage and statistically large sample will help mitigate these effects, enabling more robust, population-level trends to be established.

Furthermore, the complexity of planet formation \citep{molliere2022}, combined with our limited understanding of how metals are distributed between planetary interiors and atmospheres, as observed in the giant planets of the Solar System \citep{Miguel2022}, makes this a particularly challenging task. Without statistically validated trends across a diverse population of exoplanets, even the most advanced chemical and formation models remain constrained in their predictive power. Consequently, the identification and refinement of these fundamental trends are essential for bridging forward models and observational data, and will be a key scientific objective for upcoming missions like Ariel.

\subsubsection{Strategy for observations}
To robustly characterize the relationship between atmospheric composition and planetary properties, Ariel must implement an observation strategy that balances statistical breadth with chemical precision. This approach is critical for enabling comparative atmospheric studies across diverse exoplanet populations, as demonstrated by previous population-level analyses \citep{Sing2016,Tsiaras2018,Changeat2022,Edwards2023}.

The strategy should prioritize:
\begin{itemize}
    \item Target selection spanning a wide range of planetary masses (2--300~$M_\oplus$) and radii (1--15~$R_\oplus$)
    \item Coverage across equilibrium temperatures from $\sim$300 to 2000~K, probing key chemical regimes (e.g., water-dominated, methane-rich, CO/CO$_2$-dominated)
    \item Inclusion of host stars from spectral types F to M and a range of metallicities to assess environmental effects on composition
    \item Focusing on targets with sufficient brightness and atmospheric scale heights to reach SNR~$>$~15 for key molecular detections, enabling precise abundance retrievals
\end{itemize}

We propose observing a statistically representative subsample of approximately 100 exoplanets, including:
\begin{itemize}
    \item Hot Jupiters and warm Neptunes with extended atmospheres conducive to transmission spectroscopy
    \item Sub-Neptunes and super-Earths that probe the transition between rocky and volatile-rich planets
    \item Bright host stars ($J <$ 9), ensuring high spectral precision for atmospheric retrievals
\end{itemize}

To reach the precision required for robust constraints on quantities such as atmospheric metallicity and C/O, multiple visits per target are generally necessary. Ariel end-to-end simulations and design reference mission studies indicate that Tier 2-3 chemical characterization typically requires several visits (often 3-10 transits or eclipses, depending on system properties and cloud opacity) to achieve the SNR needed for precise abundance retrievals \citep[e.g.][]{Tinetti2018,Changeat2020,Barstow2022}. The exact number depends on stellar brightness, atmospheric scale height, and the presence of clouds or hazes.

The expected precision on retrieved metallicity and C/O depends on spectral quality, atmospheric complexity, and the presence of clouds or hazes. Dedicated retrieval studies based on simulated Ariel spectra indicate that, for clear or moderately cloudy atmospheres with strong molecular features and high S/N, uncertainties on log(metallicity) are typically $\lesssim$0.25-0.5 dex, while C/O ratios can be constrained to $\lesssim$0.1-0.3 for representative warm and hot planets with multiple carbon- and oxygen-bearing species detected across Ariel's wavelength range \citep[e.g.][]{Wang2023,Bocchieri2025}.

However, the presence of high-altitude clouds or hazes can significantly degrade these constraints. Simulations incorporating grey cloud decks show that metallicity uncertainties may increase by up to a factor of $\sim$2 when spectral features are muted, with comparable degradation for C/O as molecular signatures weaken \citep[e.g.][]{Bocchieri2025}. Furthermore, end-to-end Ariel retrieval challenge exercises demonstrate that while independent retrieval codes recover broadly consistent parameter posteriors, differences in model assumptions (e.g., thermal structure parameterization, cloud treatment, chemical prescriptions) introduce additional systematic uncertainties beyond formal statistical error bars \citep[e.g.][]{Barstow2022}. These systematic effects are expected to contribute at the level of a few tenths of a dex for metallicity and similar magnitude for C/O in typical cases.

Importantly, independent design-reference mission simulations show that even when individual targets have moderate uncertainties, a sample of several hundred planets enables statistically significant detection of population-level trends, including mass-metallicity relationships and compositional correlations with equilibrium temperature or host star properties \citep[e.g.][]{Zellem2019}. Based on these studies, a minimum of $\sim$50 well-characterized planets is sufficient to begin identifying statistically meaningful trends \citep{Sun2024}, while a larger subsample of $\sim$100 high-quality atmospheres would provide robust constraints across parameter space.

The precise list of targets will be determined in synergy with the Ariel scheduling working group, taking into account mission constraints and prioritization metrics. The estimated observational time required for this strategy, including repeated visits for key targets, remains within the nominal mission timeline and resource envelope.

Overall, while clouds, hazes, and model assumptions will increase uncertainties on a case-by-case basis, Ariel's broad simultaneous wavelength coverage and large, homogeneous sample are expected to deliver metallicity constraints typically $\lesssim$0.5 dex and C/O uncertainties $\lesssim$0.3 for favorable targets. This precision is sufficient to reveal and quantify population-level compositional trends, enabling Ariel to deliver the first compositionally resolved, statistically robust atlas of exoplanet atmospheres and to significantly advance our understanding of planet formation and evolution.

\subsection{In Which Atmospheres is Disequilibrium Chemistry Observed?}\label{sec:disequilibrium}

\subsubsection{Context}

Disequilibrium chemistry occurs when physical or chemical processes in a planetary atmosphere act on timescales that are shorter than those needed to restore the chemical composition to its state of minimal Gibbs free energy. Examples of such processes are vertical mixing, which may lead to quenching \citep{Moses2011, Venot2012, Miguel2014, Mukherjee2025}, photochemistry causing molecular dissociation and the subsequent production of long-lived species \citep[][Sect. \ref{sec:flares}]{Miguel2015, Hu2021, Baeyens2022, Konings2022}, upper-atmospheric heating, ionization, and the production of photoelectrons through X-ray and UV-absorption \citep{Locci2022, GarciaMunoz2025}, and global circulation homogenizing the composition for regions with a different temperature \citep[][and Sect. \ref{sec:circulation}]{Drummond2018_HD209458b, Baeyens2021, Zamyatina2023}. To predict whether disequilibrium processes will strongly affect the atmospheric composition, accurate knowledge of chemical pathways and the rate-limiting steps is essential \citep{Tsai2018}. As such, disequilibrium chemistry highlights the tight link between physics and chemistry in planetary atmospheres.

Although several studies have found evidence for disequilibrium chemistry on the level of individual planets \citep[e.g.][]{Tsai2023b, Bell2024} and even planet populations \citep{Baxter2021, Roudier2021}, strong trends in disequilibrium chemistry are yet to emerge. A major obstacle lies in the large planetary diversity and potential scatter in chemical properties. The Ariel space mission, with its focus on population studies of exoplanets atmospheres, provides an excellent opportunity to uncover these underlying trends.

One particular trend that is still eluding explanation is that of \textit{methane quenching}. Indeed, while planets cooler than $\sim 1000$~K are expected to have methane (CH$_4$) in their spectrum according to chemical calculations, its detection has been long overdue \citep[e.g.][]{Morley2017, Kreidberg2018, Benneke2019_GJ3470b, Carone2021, Fu2022, Barat2024, Bell2024, Dyrek2024, Zhang2025}. Nonetheless, methane has been found in a select few exoplanets \citep{Bell2023, Madhusudhan2023, Beatty2024, Benneke2024, Welbanks2024}, leading to ad-hoc disequilibrium chemistry to explain why CH$_4$ appears on some planets but not in others.
Besides its implications for the C/O ratio (see Sect.~\ref{sec:planet_formation}), an improved understanding of methane quenching may enable constraints on the interior properties of exoplanets \citep[e.g.][]{Fortney2020, Sing2024, Welbanks2024}.
It is also worth noting that the majority of exoplanets currently accessible to atmospheric characterisation are hot Jupiters, for which CO is expected to dominate over CH$_4$ due to their high temperatures. This observational bias further complicates the interpretation of methane quenching across different classes of planets.

\subsubsection{Strategy for observations}

Whether disequilibrium chemistry has a strong influence on the planetary atmosphere depends mostly on temperature and stellar high-energy irradiation. One strategy is to observe a diverse population of transiting exoplanets in terms of size and temperature around different stars, with excellent metrics for atmospheric characterization. 
As an illustrative exercise, we construct a representative subsample from the Mission Candidate Sample, consisting of planets that would reach Tier 2-quality spectra within approximately ten transits, and spanning a broad range of temperatures and sizes. Tier 2 is the level within the Ariel observing strategy at which population studies of exoplanet atmospheric compositions become achievable \citep{Tinetti2021}. This illustrative sample includes ten close-in, gaseous planets with relevant parameters, listed in Table~\ref{tab:sample1}. Their equilibrium temperature ranges from 600~K to 1600~K, focusing on transitions in carbon and nitrogen chemistry. We omit ultra-hot planets ($T_\textrm{eq} > 2000$~K), as their atmospheres require a more specialized physical treatment owing to the increasing importance of thermal dissociation, photo- and thermal ionization, and NLTE effects \citep{Helling2023, Fossati2018}. Moreover, the chemical network adopted in this part of the work describes neutral chemistry and does not include ion chemistry. We therefore focus on the temperature regime where transitions in neutral carbon and nitrogen chemistry can be investigated within a consistent modelling framework. We would therefore like to emphasize that the development of robust ion-neutral chemical schemes for both cold and (ultra-)hot planets is an important aim for future exoplanet studies.

\begin{table*}
\caption{\label{tab:sample1}Exoplanet sample and the input parameters that have been used to model them. Clarification of the symbols: the planetary equilibrium temperature (T$_\mathrm{eq}$), the planetary intrinsic temperature ($T_\mathrm{int}$), planet radius ($R_\mathrm{p}$), planet mass ($M_\mathrm{p}$), surface gravity ($g$), semi-major axis ($a$), and the effective temperature ($T_\ast$) and radius ($R_\ast$) of the host star. Stellar and planetary parameters were taken from the \textit{Ariel Mission Candidate Sample} \citep{Edwards2022}, with primary references as indicated below.}     
\centering          
\begin{tabular}{l r r r r r r r r c r}
\hline\hline  
Name & $T_\mathrm{eq}$ (K) & $T_\mathrm{int}$ (K) & $R_\mathrm{p}$ ($R_{\oplus}$) & $M_\mathrm{p}$ ($M_{\oplus}$) & $g$ (m/s$^2$) & $a$ (AU) & $T_{\ast}$ (K) & $R_{\ast}$ ($R_{\odot}$) & Transits \\ 
\hline                    
   AU~Mic~b     &  626 & 100 &  3.98 &  17.00 & 10.5 & 0.065 & 3700 & 0.75 & 2 & [1] \\
   TOI-1130~c   &  659 & 100 & 16.46 & 309.55 & 11.2 & 0.071 & 4250 & 0.69 & 2 & [2] \\
   WASP-107~b   &  720 & 100 & 10.31 &  30.51 &  2.8 & 0.055 & 4425 & 0.67 & 1 & [3], [4] \\
   HAT-P-11~b   &  802 & 150 &  4.27 &  28.60 & 15.4 & 0.053 & 4780 & 0.68 & 9 & [5], [6] \\
   HAT-P-12~b   &  926 & 150 & 10.52 &  67.06 &  5.9 & 0.038 & 4650 & 0.70 & 3 & [7] \\
   WASP-69~b    &  929 & 150 & 12.18 &  82.63 &  5.5 & 0.048 & 4700 & 0.86 & 1 & [8], [9] \\
   WASP-39~b    & 1084 & 200 & 13.94 &  89.31 &  4.9 & 0.049 & 5400 & 0.90 & 2 & [10] \\
   WASP-96~b    & 1243 & 350 & 13.17 & 155.73 &  8.8 & 0.045 & 5540 & 1.05 & 26 & [11] \\
   WASP-127~b   & 1353 & 350 & 14.39 &  52.34 &  2.5 & 0.050 & 5620 & 1.33 & 1 & [12], [13] \\
   KELT-11~b    & 1647 & 500 & 14.81 &  54.35 &  2.4 & 0.062 & 5375 & 2.69 & 1 & [14], [15] \\
\hline                  
\end{tabular}\\
\begin{minipage}{15cm}
\footnotesize{[1] \citet{Plavchan2020}, [2] \citet{Huang2020}, [3] \citet{Anderson2017}, [4] \citet{Piaulet2021}, [5] \citet{Bakos2010}, [6] \citet{Yee2018}, [7] \citet{Hartman2009}, [8] \citet{Anderson2014}, [9] \citet{Stassun2017}, [10] \citet{Faedi2011}, [11] \citet{Hellier2014}, [12] \citet{Lam2017}, [13] \citet{Seidel2020_WASP-127b}, [14] \citet{Pepper2017}, [15] \citet{Beatty_KELT-11b}}
\end{minipage}
\end{table*}
The capability of Ariel to distinguish between equilibrium and disequilibrium chemistry is assessed using forward models. 
We model the chemical composition of each planet in the sample to first order using a photochemical kinetics model \citep{Agundez2014, Baeyens2021, Baeyens2022} and chemical network \citep{Venot2020_network}. We use a one-dimensional implementation of the code and assume a solar elemental composition. It is worth noting that sulfur-bearing species are not included in this chemical network. Although sulfur-bearing molecules such as \ce{SO2} can be excellent indicators of photochemistry \citep{tsai23, Dyrek2024}, the role of sulfur in exoplanet atmospheric chemistry is not yet fully understood \citep[e.g.][]{Konings2025, Veillet2025_aa}. Additionally, spectrally active molecules such as \ce{SO2} and \ce{CS2} are expected to become observable at high metallicities, which are not explored in this experiment.  Therefore, a dedicated discussion of sulfur in exoplanet atmospheres is provided in Sect. \ref{sec:sulfur}. Temperature profiles were computed with \textit{petitCODE} \citep{Molliere2015, Molliere2017}, using the parameters listed in Table~\ref{tab:sample1}. 

Finally, model transit spectra were generated for cases with equilibrium chemistry and full chemical kinetics using \textit{petitRADTRANS} \citep{Molliere2019} and the Ariel noise generator \textit{ArielRad} \citep{Mugnai2020}.  As such, Ariel's ability to distinguish between the two models can be estimated based on the 1$\sigma$-error bars of the equilibrium and kinetic model data. Please note that, for clarity reasons, we did not apply an additional resampling to scatter the data points with respect to the calculated model spectrum. A measured spectrum would have such scatter. The results are shown in Figures \ref{fig:chem_spec_cool} and \ref{fig:chem_spec_hot}.

It is clear that chemical transitions occur as planets increase in equilibrium temperature. Figs.~\ref{fig:chem_spec_cool} and \ref{fig:chem_spec_hot} show \ce{CO} replacing CH$_4$ as the dominant carbon-bearing species as temperature increases. At the same time, chemical disequilibrium caused by vertical mixing and photochemistry can be seen in the concentrations of \ce{HCN} and \ce{CO2}. In many model spectra, these chemical changes can be seen in the absorption bands of methane (3.3~$\mu$m) and carbon dioxide (4.5~$\mu$m). In the coldest planets, NH$_3$ quenching forms the main difference between the equilibrium and full chemical kinetics spectrum. Absorption bands of HCN -- while being a great indicator of photochemistry -- unfortunately strongly overlap with \ce{CH4} and \ce{H2O}, making it difficult to observe this molecule unambiguously. The addition of another, more easily identifiable tracer of photochemistry such as \ce{SO2} \citep{Tsai2023b} may resolve this issue (see Sect. \ref{sec:sulfur}). This would, however, necessitate a broader parameter study, as the observability of \ce{SO2} has been linked to metallicity, gravity, $K_{zz}$, and the spectral energy distribution of the host star \citep{Konings2025}.

\begin{figure}
    \centering
    \includegraphics[width=\linewidth]{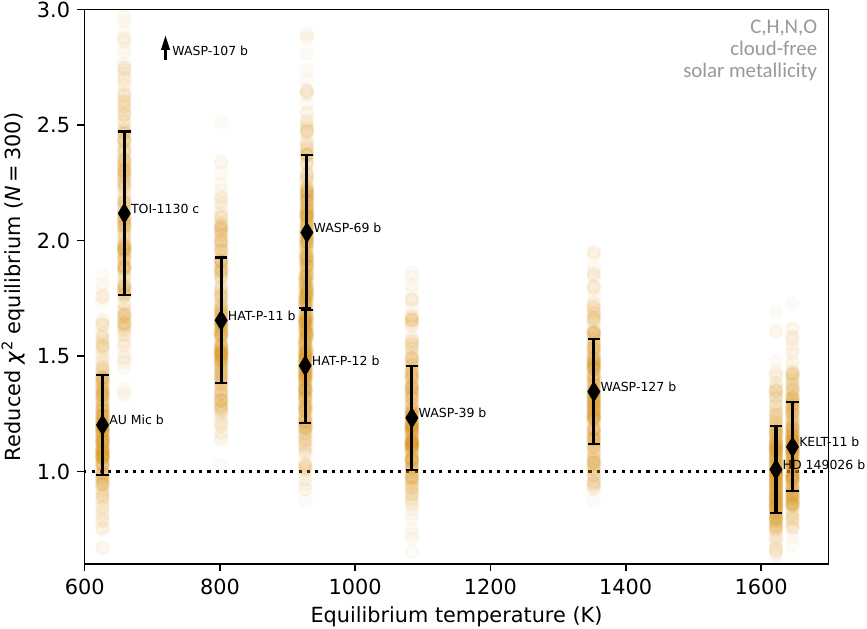}
    \caption{Reduced chi-squared values for chemical equilibrium model fits to photochemical mock observations. Black data points show the mean and standard deviation of 300 randomly fitted spectra per planet (\textit{orange circles}).}
    \label{fig:reduced_chisquare}
\end{figure}

\begin{figure*}
 \centering
  \includegraphics[width=0.90\textwidth]{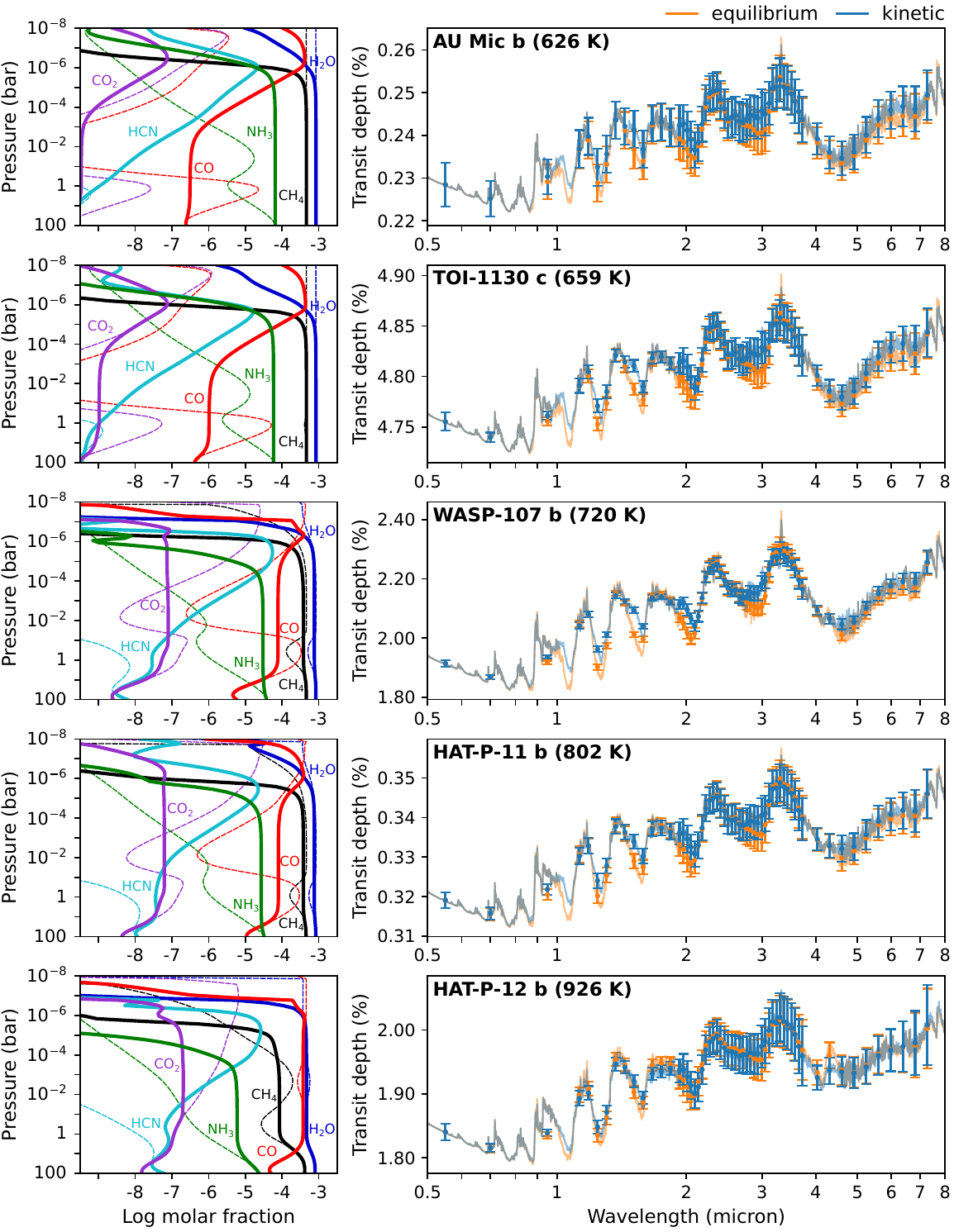}
  \caption{\label{fig:chem_spec_cool}Forward models for the coolest five planets in our sample, ordered by increasing equilibrium temperature, showing the chemical abundances (\textit{left}) in chemical equilibrium (\textit{dashed}) and full chemical kinetics (\textit{full}), and the synthetic transmission spectra (\textit{right}) for the equilibrium (\textit{orange}) and kinetics (\textit{blue}) case. For the computation of the Ariel uncertainties we assumed a Tier 2-scenario.}
\end{figure*}

\begin{figure*}
  \centering
  \includegraphics[width=0.90\textwidth]{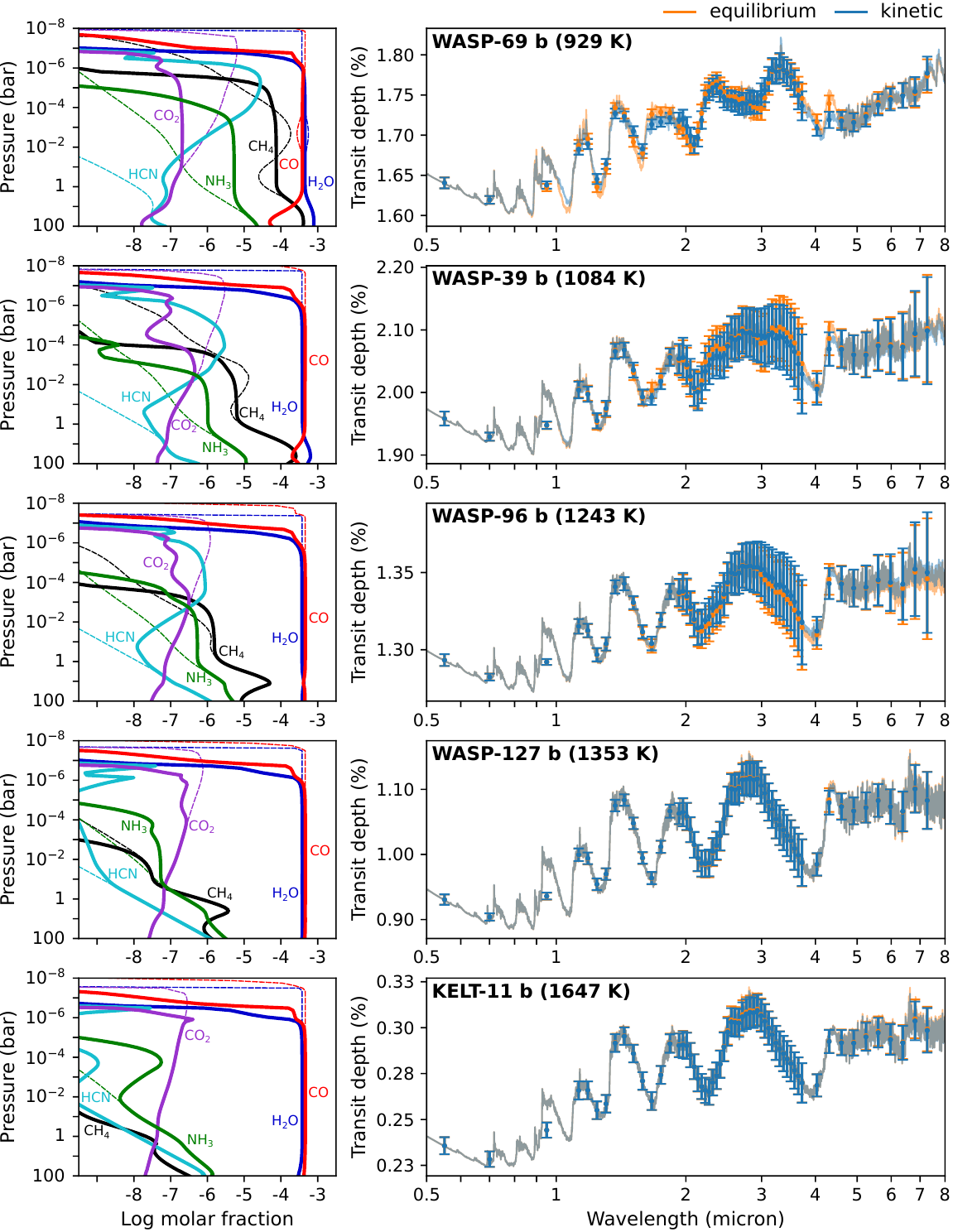}
  \caption{\label{fig:chem_spec_hot}Forward models for the hottest five planets in our sample, ordered by increasing equilibrium temperature, showing the chemical abundances (\textit{left}) in chemical equilibrium (\textit{dashed}) and full chemical kinetics (\textit{full}), and the synthetic transmission spectra (\textit{right}) for the equilibrium (\textit{orange}) and kinetics (\textit{blue}) case. For the computation of the Ariel uncertainties we assumed a Tier 2-scenario.}
\end{figure*}

Crucially, the broad wavelength coverage of Ariel will be able to clearly distinguish between water and methane absorption, something which has been difficult with narrow wavelength coverage \citep{Bezard2020}. Although overlapping spectral features pose challenges, retrieval simulations indicate that Ariel has nonetheless the potential to constrain key molecules such as HCN, \ce{CH4}, and \ce{H2S}, and retrieving elemental ratios (C/O, N/O, S/O) in hot-Jupiter atmospheres \citep{wang2023_constraining}.
Here, we make a first order assessment of Ariel's capability to discern trends in disequilibrium chemistry as a function of planetary equilibrium temperature. To do so, we fit chemical equilibrium models to the -- more realistic -- photochemical kinetics spectra. Each data point has random scatter applied to it, in accordance with the calculated Ariel uncertainties. Finally, we calculated reduced chi-squared values of the chemical equilibrium fit to the randomly generated mock photochemistry spectra for each planet. This experiment was done 300 times and the resulting values, along with their mean and standard deviation are shown in Fig.~\ref{fig:reduced_chisquare}. The data show that planets in the [650-1000]~K regime statistically deviate from chemical equilibrium ($\chi^2_\textrm{red} = 1$). Planets with $T_\textrm{eq} > 1000$~K appear to be well fit by the chemical equilibrium model, with $\chi^2_\textrm{red}$-values closer to one as a result. WASP-107~b is an outlier ($\chi^2_\textrm{red}\approx 7$) and, with its very precise spectrum, represents one of the best targets to detect disequilibrium chemistry \citep[see also][]{Dyrek2024, Welbanks2024, Sing2024}.

We stress that the analysis above represents an idealized situation in which the atmosphere is assumed to be free of clouds. Nonetheless, clouds are ubiquitous in planetary atmospheres \citep{Sing2016}, and while trends with temperature have been established \citep[e.g.][]{Gao2020, Helling2023}, it is not yet understood why some atmospheres appear cloudier while other are relatively cloud-free. If present, clouds can interfere with atmospheric retrievals by muting spectral features in a way that is degenerate with chemical abundance \citep{Barstow2017, Pinhas2019, welbanks19}. Thus, trends such as the one shown in Fig.~\ref{fig:reduced_chisquare} will have more intrinsic scatter if some atmospheres have a (partial) cloud cover, and larger samples will be needed to infer the underlying properties of the population. This issue will be addressed in Ariel's observing strategy, with $\sim$600 planets in the Tier-2 chemical survey \citep{Edwards2022}.

Finally, we note that free parameters such as the eddy diffusion coefficient, atmospheric metallicity, C/O, and global redistribution \citep[][and Sect. \ref{sec:circulation}]{Venot2014, Venot2020_wasp43b, Drummond2020, Zamyatina2023} will complicate the interpretation of atmospheric chemistry and which atmospheres will be out-of-equilibrium \citep{Miguel2014}. As an example, the combination of strong interior heating with vigorous vertical mixing results in a depletion of methane on WASP-107~b as compared to our model \citep{Welbanks2024, Sing2024}. Additionally, global circulation linking the hot day-side and cold night-side of planets can lead to chemical disequilibrium even in ultra-hot planets \citep{Baeyens2024}. As such, we advocate for a large and diverse sample of transit spectroscopy observations (beyond the candidate sample listed in Table~\ref{tab:sample1}), including phase curve observations to break degeneracies as well as maximal synergies with the James Webb Space Telescope \citep{charnay2022_ExAstro, changeat2025_synergy}.

\subsection{How does atmospheric circulation affect atmospheric chemistry?}\label{sec:circulation}

\subsubsection{Context}

Advection is the transport of air together with its constituents and properties by the bulk atmospheric flow. Advection does not chemically transform the transported air itself, but by moving it through different regions of the atmosphere drastically changes where and how chemical species are produced, transformed, or destroyed, and alters the effective lifetime of chemical species.  Contrary to diffusion that is driven by random atmospheric motions, advection is driven by the large-scale atmospheric circulation, and acts over large distances. All that together means that advection leaves signatures of atmospheric circulation in the chemical composition of an atmosphere.

Advection transports air downwind. To understand advection, it is common to break it down into three components: zonal (east-west, or across longitudes), meridional (north-south, or across latitudes), and vertical (upward or downward) advection. That helps identify the predominant direction of advection, and show that, for example, zonal advection dominates inside jet streams, or that meridional and vertical advection control the equator-to-pole overturning circulation. Horizontal advection is usually the sum of zonal and meridional advection.

Numerical models that always explicitly include advection are called General Circulation Models (or GCMs). GCMs that couple atmospheric dynamics, radiative transfer and disequilibrium chemistry (see Sect.~\ref{sec:disequilibrium}) are often called Chemistry-Climate Models (or CCMs). CCMs were applied to study atmospheres of a variety of exoplanets, from cold rocky planets (e.g., \textit{WACCM}, \citealt{Chen2019}; \textit{UM-UKCA}, \citealt{Braam22_lightninginduced}; \textit{UM}, \citealt{Ridgway22_3d}; and \textit{PCM}, \citealt{Stolzenbach23_threedimensional}) to hot gas giants (e.g., \textit{SPARC-MITgcm}, \citealt{Cooper2006}; \textit{THOR} \citealt{Mendonca18_threedimensional}; \textit{UM}, \citealt{Drummond2018_HD189733b,Drummond2018_HD209458b,Drummond2020}; and \textit{Exo-FMS}, \citealt{Lee2023a,Lee24_dynamically}). CCM simulations show that advection affects atmospheric chemistry differently for different planets as they host different atmospheric circulations.

For tidally-locked \ce{H2}-dominated gas giants, which are the primary targets of Ariel, CCMs generally predict that advection homogenises abundances of gas-phase chemical species in the observable atmosphere by smoothing out their abundance gradients otherwise expected at chemical equilibrium. Long-lived chemical species like \ce{CH4}, \ce{NH3}, and \ce{HCN} are most affected, and to a lesser extent \ce{CO2} \citep[e.g.,][]{Lee2023a,Zamyatina2023}. This homogenisation is caused by transport-induced quenching, which is a process that occurs when advection (but also, more generally, mixing via advection and/or diffusion) prevents the establishment of chemical equilibrium by acting on a shorter timescale than the chemical timescale. Both horizontal and vertical advection affect quenching. However, CCMs disagree on their relative importance, and hence on the exact homogenised atmospheric abundances. For example, \citet{Cooper2006} found that vertical advection controls quenching for HD~209458~b, while \citet{Mendonca18_threedimensional} and \citet{Lee2023a} showed that zonal advection prevails for WASP-39~b, WASP-43~b, and HD~189733~b. The role of meridional advection was highlighted in \citet{Drummond2018_HD189733b}, \citet{Drummond2018_HD209458b}, and \citet{Drummond2020}, and the dependence of quench level on latitude was discussed in \citet{Lee2023a} and \citet{Zamyatina24_quenchingdriven}.

To date, no published CCM used to study atmospheres of gas giants included photochemistry. However, photochemistry is important for this type of planets (see Sects.~\ref{sec:disequilibrium}, \ref{sec:sulfur}, and \ref{sec:flares}). The best available framework capable of estimating the impact of zonal advection on atmospheric photochemistry of gas giants is \textit{VULCAN 2D} \citep{Tsai24_global}. This model considers zonal advection together with zonal and vertical diffusion, with advection transporting air in an equatorial plane from the upwind cell only. \textit{VULCAN 2D} simulations showed that photochemical products are advected from the permanently irradiated dayside to the nightside. These products alter the nightside chemistry, and cause, e.g., \ce{SO2} accumulation and \ce{HCN} and \ce{C2H4} production on the nightside \citep{Tsai2023c,Tsai24_global}.

\subsubsection{Strategy for observation}

\begin{figure*}
    \centering
    \includegraphics[width=0.95\textwidth]{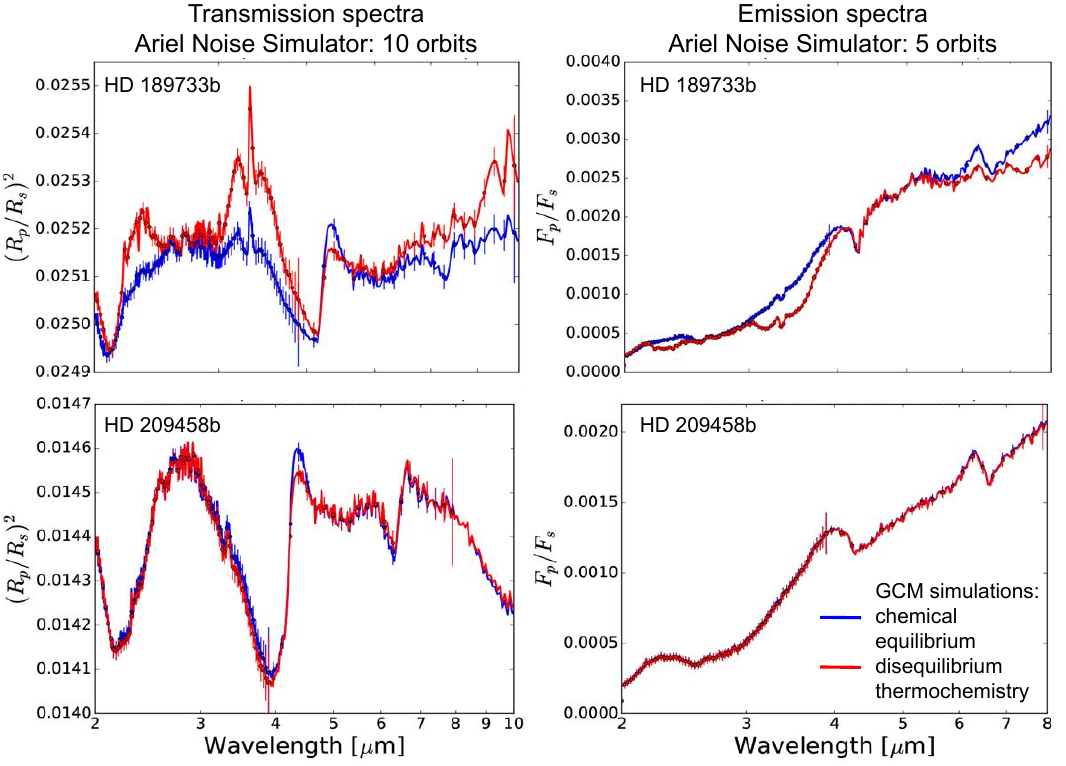}
    \caption{Detectability of signatures of transport-induced quenching in the terminator-average transmission and emission spectra of HD~189733~b and HD~209458~b according to the Ariel noise generator \textit{ArielRad} \citep{Mugnai2020} applied to the output from the Met Office \textit{Unified Model}, configured as in \citet{Drummond2020} assuming an aerosol-free atmosphere of solar \citep{Asplund2009} composition. Adapted from \citet{Drummond2020}. Signatures of transport-induced quenching would be more detectable in atmospheres of HD~189733~b-like planets.}
    \label{fig:Drummond20_ariel}
\end{figure*}

Systematic exploration of the impact of atmospheric circulation on atmospheric chemistry of exoplanets is an ongoing effort. There are, however, studies that explored that impact for individual planets. Studies using \textit{VULCAN 2D} showed that for hot Jupiters HD~189733~b, HD~209458~b, and WASP-39~b zonal advection prevails over vertical mixing at pressures higher than 0.1 mbar, while at lower pressures photochemistry and vertical mixing control the composition \citep{Tsai2023c,Tsai24_global}. Studies using the Met Office \textit{Unified Model (UM)} suggest that amongst the following list of warm and hot Jupiters, HAT-P-11~b, HD~189733~b, HD~209458~b, WASP-15~b, WASP-17~b, WASP-39~b, and WASP-96~b \citep{Zamyatina2023,Zamyatina24_quenchingdriven,Espinoza24_inhomogeneous,Kirk25_bowiealign}, HD~189733~b and WASP-96b are at the ``sweet spot'' for observability of signatures of transport-induced quenching via a detection or equally a non-detection of \ce{CH4} at $3.3 \mu m$ in the terminator-average transmission and emission spectra (Figure~\ref{fig:Drummond20_ariel}). A detection of \ce{CH4} would suggest that atmospheric composition of these planets is close to solar, as it would enable \ce{CH4} to be equally abundant throughout the observable atmosphere thanks to transport-induced quenching.
A non-detection of \ce{CH4}, however, would imply atmospheric compositions higher than solar, which would cause \ce{CH4} to get depleted by quenching instead \citep[see][Figure 7 for WASP-96b]{Zamyatina24_quenchingdriven}. Given this limited evidence and the fact that the aforementioned studies into impacts of atmospheric circulation on chemistry assume aerosol-free atmospheres, the recommendation for Ariel for strategy for observation of signatures of large scale atmospheric circulation ``through the lens'' of gas-phase only chemistry is to target HD~189733~b-like hot Jupiters.

\subsection{How important is sulfur and phosphorous chemistry? In which kind of atmosphere these elements are abundant?}\label{sec:sulfur}

\subsubsection{Context}
After hydrogen, helium, oxygen, carbon, and nitrogen, sulfur and phosphorus are the two next non-metallic and gaseous elements in terms of abundance in the Universe. Their presence in planetary atmospheres has been thus anticipated even before their first detection. 
Sulfur is common in the solar system, and it is present in the form of H$_2$S in giant planets and icy giants \citep{niemann1996, atreya1999,Irwin2018}, on its oxidized form (SO$_2$ and H$_2$SO$_4$) on the clouds of Venus \citep{mills_allen2007}, SO$_2$ frost and volcanic sulfur species on Io \citep{spencer2000, spencer2005}, and, potentially, sulfur compounds (in the form of H$_2$S) in the plumes of Enceladus \citep{waite2009}. On the other hand, phosphorus is rarer and more difficult to detect. Phosphine (PH$_3$) has been confirmed in Jupiter and Saturn \citep{Fletcher2009} due to disequilibrium chemistry (i.e., vigorous vertical mixing in gas giants), while its detection on Venus is highly controversial \citep{greaves2021,snellen2020, cordiner2022}. On Earth, phosphorus is mainly in the solid phase, as phosphates, in crustal and oceanic reservoirs \citep{filippelli2008}. All these information regarding the solar system offers a guide towards where (and in which form) sulfur and phosphorus should be expected in exoplanet atmospheres.

Concerning exoplanetary atmospheres, the presence of S- and P- species was predicted by \cite{Wang2017, Tsai2021, Polman2023}, and it has been possible to achieve the first detection of a sulfur species (not the expected one) thanks to the high sensitivity of JWST instruments \citep{Alderson2023, Rustamkulov2023}. The detection of this sulfur species, namely \ce{SO2}, has been of particular interest because its presence in such abundance has only been explained by the disequilibrium process of photochemistry \citep{Tsai2023b}. Since then, \ce{SO2} and a few other sulfur species (H$_2$S, CS$_2$, CS, and H$_2$CS) have been detected in a handful of exoplanets (HD~189733~b, \citealt{Fu2024, Zhang2025}; WASP~107~b, \citealt{Dyrek2024}; GJ~3470~b \citealt{Beatty2024}; TOI-270~d, \citealt{Benneke2024, Felix2025, Holmberg2024}; TOI-421~b, \citealt{Davenport2025}). These detections, and in particular the relative abundances between some sulfur species (CS$_2$/SO$_2$), suggest that some planets (e.g. TOI-270 d) could exhibit an atmospheric sulfur enrichment compared to that of the parent star \citep{Felix2025}. Knowing the amount of sulfur in exoplanet atmosphere is of particular interest as it can be a footprint of the formation history of those planets \citep{Kama2019, Turrini2021, Pacetti2022, Crossfield2023}. Thus, understanding the dynamical and chemical processes that can influence and control the abundances of sulfur species is crucial.
In this view, several theoretical work \citep{Polman2023, veillet2025_pccp} and modeling studies \citep{Tsai2023b, Tsai2023c, moses2024DPS, Veillet2025_aa} have been performed recently. In brief, in warm exoplanet atmospheres, it is expected than H$_2$S, the equilibrium sulfur species is destroyed by photodissociations and/or reactions with H. However, various sulfur molecules can emerge across temperatures, due to the wide redox state of sulfur. Understanding the sulfur reservoir is essential in constraining metallicity and formation history.
Then, depending on various parameters, such as metallicity, atmospheric temperature, irradiation, different species can be present in detectable amounts: for instance, CS$_2$ is expected in atmospheres in high metallicity conditions ($\geq$ 100$\times$solar) with an equilibrium temperature in the [270-720] K range (with a peak centered around 570 K). In contrast, SO$_2$ is expected at higher equilibrium temperatures (T$_\mathrm{eq}$ $>$ 900 K). It has also been found/shown that the presence of NH$_3$ could play an important role in sulfur chemistry, leading to the formation of NS \citep{Tsai2021}. Finally, it is important to note that some experimental works \citep{Reed_2024} have confirmed that several S-species, such as CS$_2$, are photochemical products in relevant conditions.\\

Concerning phosphorus, it was first thought that phosphine signature would be the only species visible at 4-5 $\mu$m, especially for warm atmospheres (T$_\mathrm{eq}$=500 K and T$_\mathrm{eq} <$ 1000 K,  \citealt{Wang2016}) and the first detection of PH$_3$ in an extrasolar atmosphere \citep[i.e. brown dwarf,][]{Rowland2024} further supports the idea that this molecule is one of the most promising to be detected.
However, considering kinetic chemistry, it seems that HOPO, PO, and P$_2$ might be more likely to be detectable. In particular, the large PO abundance (photochemically produced) could be detectable at 4.1 $\mu$m and 8 $\mu$m, especially in high metallicity atmospheres \citep{Lee2024}. Only the 4.1 $\mu$m feature is really relevant for Ariel, but unfortunately, this one could be also obscured by SO$_2$ feature. However, it is important to keep in mind that the phosphorus models used for these predictions are still uncertain due to poor knowledge on P chemistry \citep[e.g.][]{Wang2023, Zilinskas2025}, and lack of some phosphorus species line-lists. New chemical schemes are currently being developed (Le Cadre, Venot et al. in prep) that could lead to different results. For instance, other P-species could be more abundant than predicted up to now and thus detectable.  

Ariel, with its observation strategy, is ideal for meeting the current need. We will be able to test our current predictions and better understand the parameters that govern sulfur and phosphorous chemistry in hot exoplanets.

\subsubsection{Strategy for observations}

In order to detect sulfur and phosphorous species, we recommend both transmission and/or emission spectroscopy.

\paragraph{For phosphorous:}
The first detection of phosphine in an extrasolar atmosphere (a brown dwarf; \citealt{Rowland2024}) highlights the potential for detecting phosphorus-bearing species in exoplanet atmospheres. Informed by recent developments in photochemical modeling (PHO network, \citealt{Lee2024}), we propose a targeted strategy based on atmospheric properties. Cold, solar-metallicity planets (T$_\mathrm{eq}$$\sim$300-600 K) with limited photochemical activity remain the most favorable environments for PH$_3$ retention and potential detection \citep{Mukherjee2022, Rowland2024}. However, PH$_3$ is unlikely to be observable in warmer or more irradiated atmospheres, where photochemical processing efficiently destroys it. In contrast, warmer planets (T$_\mathrm{eq}$$\sim$600-1500 K), especially those with enhanced metallicity, are expected to host detectable levels of other P-bearing species such as PO and HOPO, with PO being the most promising candidate given current spectroscopic data. PO shows absorption features around 4.1 and 8 $\mu$m, which are partially accessible to Ariel. We therefore recommend prioritizing two complementary categories of targets:
\begin{itemize}
\item Cool planets with minimal UV flux to search for PH$_3$, such as GJ~3470~b or GJ~1214~b.
\item Hot, metal-rich planets where photochemistry may enhance the abundance of PO and HOPO, such as HD~189733~b, WASP-12~b, or WASP-33~b.
\end{itemize} 
This selection strategy will enable Ariel to probe the diversity of phosphorus chemistry and place meaningful constraints on the detectability of key P-bearing molecules.
However, it is important to remember that the significant uncertainties in our knowledge of phosphorus kinetics and photochemical data might affect these predictions due to these blind spots in the models, which could result in artifacts that impact the abundance of multiple species.

\paragraph{For sulfur:}
Given the recent detections of sulfur species (e.g., SO$_2$, CS$_2$) in exoplanetary atmospheres -- particularly in systems like TOI-270 d, where CS$_2$/SO$_2$ ratios suggest sulfur enrichment \citep{Felix2025}-- and the theoretical predictions highlighting temperature- and metallicity-dependent reservoirs \citep{Veillet2025_aa}, we recommend a broad observational strategy. Despite these advances, uncertainties in photochemical pathways and spectroscopic data persist \citep{Huang_2024}, motivating our inclusion of a diverse range of planet types to maximize detection opportunities. We therefore propose a two-pronged observational strategy for Ariel:
\begin{itemize}
    \item Warm to hot exoplanets with moderate to high metallicity and strong UV irradiation, where photochemical production of \ce{SO2} and \ce{CS2} is expected to be most efficient. Well-studied examples include HD~189733~b, WASP-39~b, TOI-421~b, TOI-270~d, HAT-P-1~b, and HD~106315~c.
    \item Planets with moderate temperatures (T$_\mathrm{eq}$$\sim$270-720 K) and very high metallicity ($>$100$\times$solar), promising for detecting \ce{CS2} and potentially other complex sulfur species, such as GJ~3470~b and HAT-P-11~b.
\end{itemize}
Ariel's broad wavelength coverage and survey strategy, especially in the 4-8 $\mu$m range where \ce{SO2} and \ce{CS2} have strong absorption features, are ideally suited to probing this chemical diversity. By systematically observing a large and varied sample of exoplanets spanning a wide range of temperatures and metallicities, Ariel will:
\begin{itemize}
    \item Track transitions between different sulfur reservoirs,
    \item Constrain sulfur abundances and the role of photochemistry,
    \item Provide insights into planet formation and atmospheric evolution through sulfur chemistry.
\end{itemize}

This observational approach complements the strategy for phosphorus species and will help build a more complete picture of volatile element cycles in exoplanet atmospheres.

\subsection{In which Exoplanet atmospheres do PAHs form? }\label{sec:PAH}

\subsubsection{Context}
Polycyclic aromatic hydrocarbons (PAHs) and related species have been found to be an omnipresent component of matter in the interstellar medium as evidenced by their abundant mid-infrared emission features (e.g., see \citealt{Hrodmarsson2025} and references therein). Typically, PAHs are composed of fused hexagonal carbon rings with hydrogen atoms attached at the peripheries, but PAHs can be formed in an infinite variety of sizes and shapes. They can be fused with hetero atoms like nitrogen, oxygen, phosphorus, and sulfur, and in the interstellar medium (ISM) several charge states are expected to be of importance, ranging from -1 for PAH anions to +2 for PAH dications \citep{Berne2022}. Having been proposed as the originators of the unidentified infrared red bands \citep{Leger1984, allamandola1985}, they have since been found to be among the most widespread organic compounds in space \citep{joblin2011pahs}. Although their IR features are detected across diverse astronomical environments such as photodissociation regions formed around massive star-forming regions, planetary nebulae, and protoplanetary disks \citep{hony2001, peeters2002, peeters2024, habart2024, chown2024}, only recently have individual PAHs been detected in the TMC-1 molecular cloud using their microwave fingerprints \citep{McGuire2021, Burkhardt2021, Cernicharo2021, Cernicharo2024, wenzel2024, wenzel2025a, wenzel2025b}. They have also been detected in Solar System bodies like Titan \citep{dinelli2013,Lopez-Puertas_2013}, in meteorite samples \citep{basile1984polycyclic, naraoka2000, naraoka2023, Oba2023}, and in Earth's atmosphere \citep{lammel_polycyclic_2015,Dat_2017}.

Their significance extends beyond astrochemistry; PAHs are also implicated in prebiotic chemistry, potentially catalyzing steps toward the origin of life by contributing to amino acid and nucleotide formation \citep{ehrenfreund2006experimentally, EHRENFREUND2007383,rapacioli2006formation,ehrenfreund2000organic,wakelam2008polycyclic,kim20123,puzzarini2017spectroscopic,closs2020prebiotically}. They also exhibit spectral degeneracy with clouds and hazes in the optical slope region \citep{grubel2025detectability,arenales2025polycyclic}.
Closer to home, PAHs play an important role as intermediates in the nascence and agglomeration of soot particles, a principal component of particulate matter created by incomplete combustion of gasoline, diesel, and coal \citep{haynes1981soot, richter2000formation}. PAHs pose a significant risk to human health as many are documented to be toxic, mutagenic, and/or carcinogenic \citep{abdel2016review}. 

PAHs all do share some commonalities in terms of their photophysical behavior. The similarities observed in their photoabsorption and photoionization cross sections have been well documented from both experiments and theoretical calculations (see \citealt{malloci_electronic_2004, hrodmarsson_photoionization_2025}). PAHs have long been invoked as contributors to the UV extinction bump which is well worth mentioning and has been reviewed in the past (See \citealt{Mulas2011}). The plasmon resonance at 17-18 eV which is universally observed in PAHs (see \citealt{hrodmarsson_photoionization_2025} and references therein), is not solely due to $\sigma$* ← $\sigma$ transitions but also $\sigma$* ← $\pi$, $\pi$* ← $\sigma$ and Rydberg spectral transitions \citep{malloci_electronic_2004}. Collectively, these lead to photoionization of PAHs that is considered the principal heating mechanism in the ISM \citep{Tielens2008}.

PAHs are usually identified via strong bands at 3.3, 6.2, 7.7, 8.6, 11.2 and 12.7 $\mu$m \citep{Tielens2008, Li2020NatAs, peeters2021spectroscopic}. However, the profiles of the main PAH bands manifest subtle diversities in relative band strengths and band profiles within individual astrophysical objects and within extended sources (see Fig. 2 in \citealt{peeters2021spectroscopic}). These can be classified into four classes, A, B, C, and D, where each is principally based on the band positions with subtle changes in intensities. The appearance of these classes of PAH bands is known to depend intricately on the object type, and hence their environment \citep{peeters2002, vanDiedenhoven2004, Matsuura2014}.  This means that although the IR signatures of PAHs appear omnipresent in various cosmic environments, detecting them in exoplanet atmospheres may come with severe complications as their principle spectral features cannot only overlap with other organic species, the spectral features of the PAHs change depending on the environment as the environment directly influences the chemistry of PAHs. 

\begin{figure*}
\centering
\includegraphics[width=0.9\linewidth]{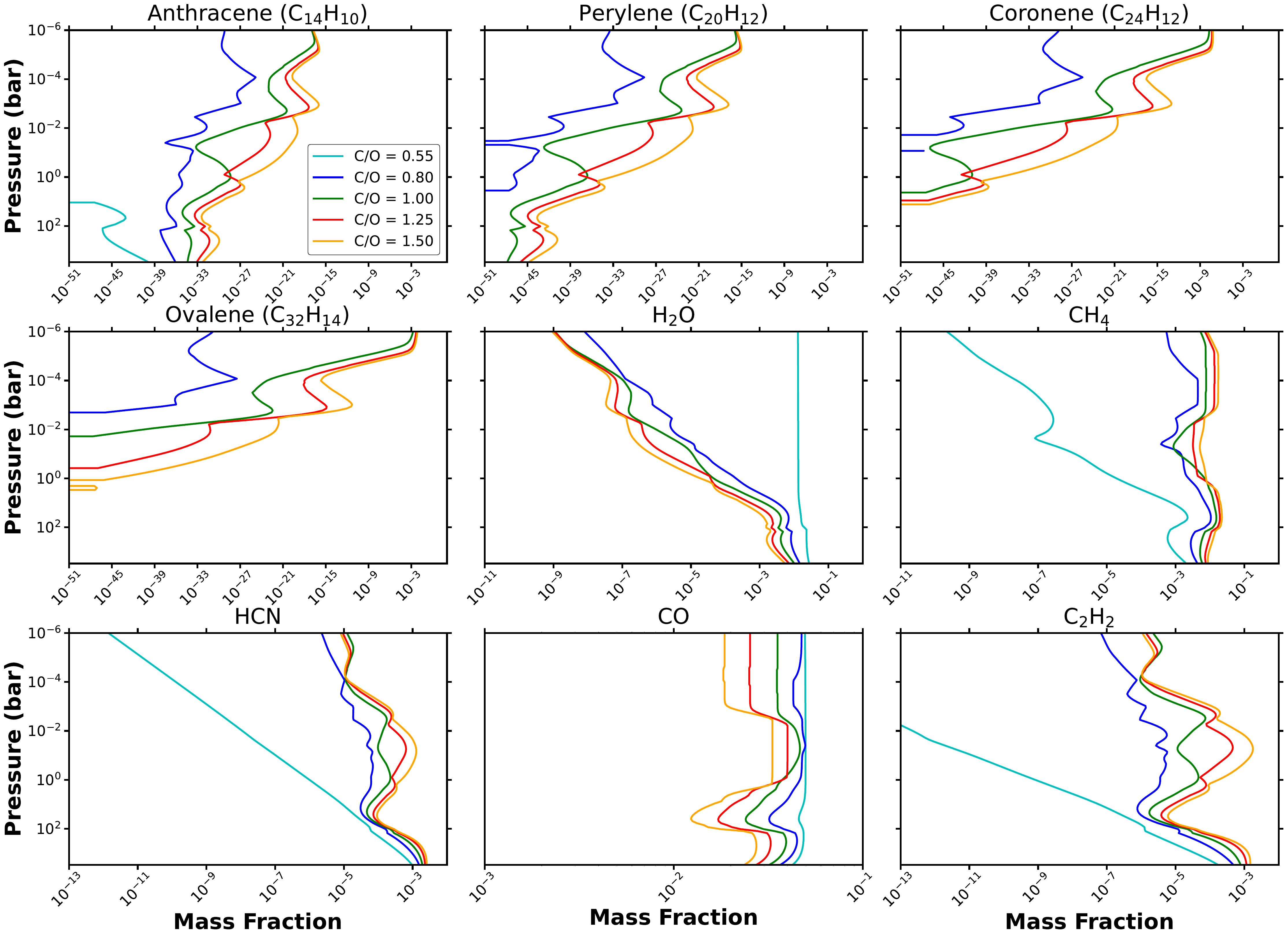}
\caption{Mass fractions of four representative PAH species alongside key molecular absorbers ($\mathrm{H_2O}$, $\mathrm{CH_4}$, HCN, CO, and $\mathrm{C_2H_2}$) in a transiting exoplanet atmosphere with $T_{\mathrm{eff}}$ = 1300K, $T_{\mathrm{int}}$ = 200K, [Fe/H] = 1.0, and $\log(g) = 3.0$, shown as a function of varying C/O ratios (taken from \citealt{dubey2023polycyclic}).}
\label{fig:VMRs}
\end{figure*}

\begin{figure}
\centering
\includegraphics[width=\linewidth]{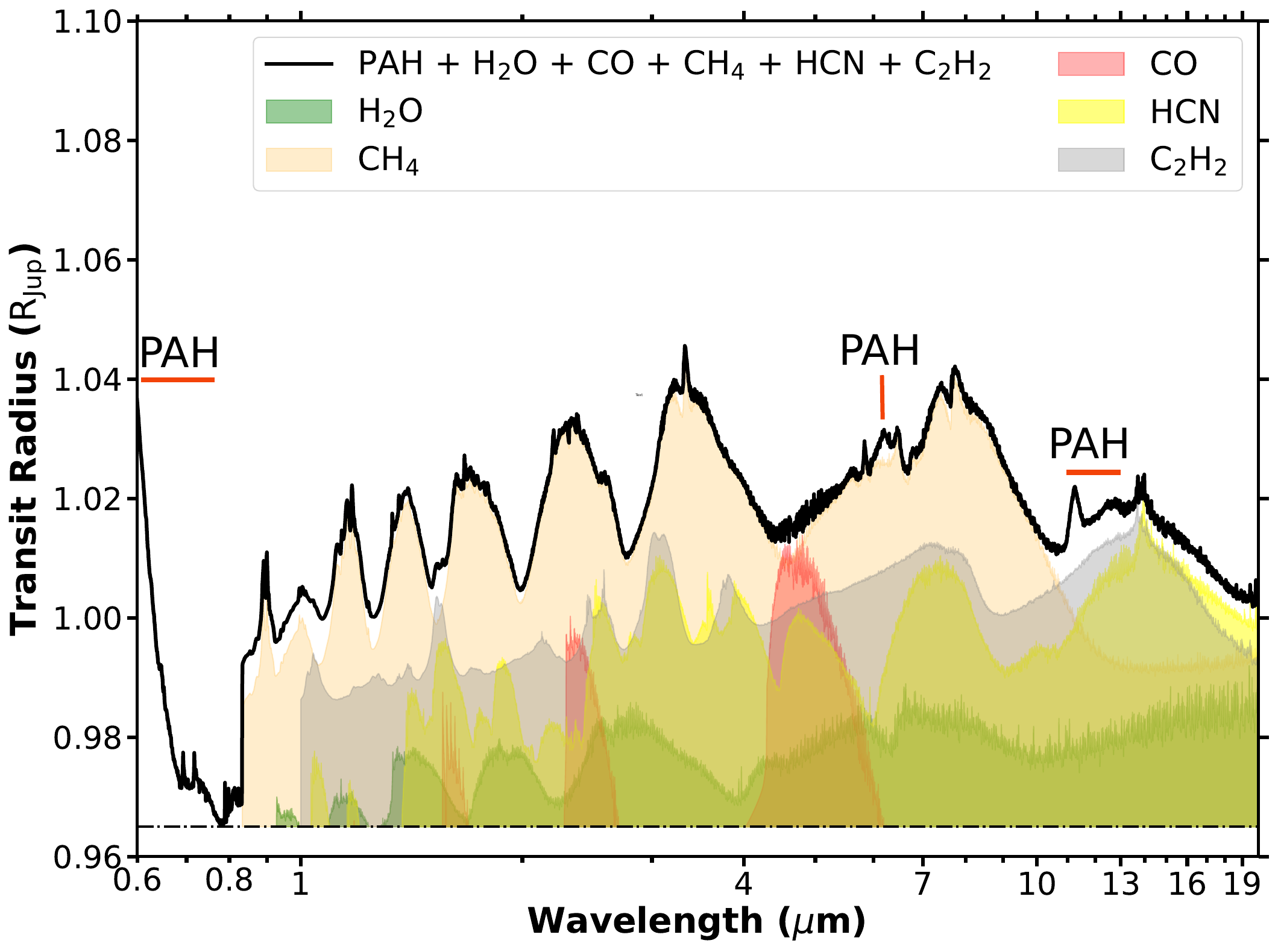}
\caption{Transmission spectrum for the synthetic planet (T$_\mathrm{eq}$~=~1300K, C/O~=~1.2, [Fe/H]~=~1.0, and log(g)~=~3.0) with key contributors to the atmospheric spectrum (taken from \citealt{dubey2023polycyclic}).}
\label{fig:contribution}
\end{figure}

Hence, PAHs have remained elusive in exoplanet atmospheres. Besides the issues raised, this gap also stems from both the lack of targeted models and the limitations of previous observational facilities.  The recent study by \cite{dubey2023polycyclic} represents a breakthrough: using thermochemical equilibrium models via the 1D radiative convective thermochemical equilibrium model, they showed that PAHs can form in thermalized atmospheres of hot Jupiters, particularly under certain conditions: equilibrium temperatures (T$_\mathrm{eq}$) around 1300K, 
elevated metallicities, and super-solar carbon-to-oxygen (C/O) ratios (see Fig.~\ref{fig:VMRs} for mixing ratio profiles and Fig.~\ref{fig:contribution} for spectral contribution of PAHs in a supersolar C/O atmosphere). The C/O ratio was found to exert the dominant influence. This is quite sensible given that PAHs tend to form in oxygen-depleted environments. On Earth, they tend to form during incomplete combustion in part due to the inherent stability of the propargyl (C$_3$H$_3$) radical which recombines to form the first aromatic ring \citep{hrodmarsson2024isomer}, while in space, PAH formation can be driven by several groups of reactions to give rise to PAH formation in cold molecular clouds, hydrogen-rich atmospheres of planets and their moons, or carbon-rich Asymptotic Giant Branch (AGB) Stars  \citep{kaiser2021aromatic}. These modeling results exhibit strong concordance with findings from laboratory experiments \citep{2018NatAs...2..303H,he2023optical} and molecular dynamics simulations \citep{Hanine_2020,hirai2021molecular}.

It is worth expanding on the likely influence of the stellar UV flux on PAH chemistry, both with respect to photochemistry as well as chemistry of fragments formed from photodissociation or photoionization. Today, there are no complete PAH chemical models that include temperature and pressure dependent information about PAH formation, destruction, and relaxation because a majority of these data are not available. Multiple reactions have been studied (see, e.g., \citealt{kaiser2021aromatic} and Tables 3 and 4 in \citealt{Hrodmarsson2025}). But typically, these are limited to singular temperature and pressure conditions. Extrapolations of measured products and yields to other temperatures and pressures require high level calculations. These are exceedingly difficult in the case of PAHs considering their size and, axiomatically, the N dimensionality of the potential energy surfaces involved in their formation and destruction.  

It is also worth mentioning that in PAHs there are complex internal molecular rearrangements after energetic processing that are only now beginning to become understood \citep{patch_radical_2025}.

Understanding whether and where PAHs form is crucial for tracing complex organic chemistry in planetary environments and has implications for atmospheric modeling, photochemistry, and even habitability assessments. Also,  considering the pivotal role PAHs play in the formation of soot and dust, they are important to interpret the cloud and haze properties inside a planet's atmosphere. However, to robustly identify the presence of PAHs given the difficulties involved, observational confirmation is required. This is something that current and next-generation instruments like Ariel are well-positioned to address.

\subsubsection{Strategy for observations}

Ariel delivers simultaneous spectroscopy from 0.5 to 7.8 $\mu$m: VISPhot (0.5-0.6 $\mu$m), FGS-1 (0.6-0.8 $\mu$m), NIRSpec (1.1-1.95 $\mu$m), AIRS-0 (1.95-3.9 $\mu$m), AIRS-1 (3.9-7.8 $\mu$m). This continuous, low-resolution (R $\approx$ 50-200) coverage captures both the optical haze slope and the suite of mid-IR PAH bands in a single or multiple visits. Particular attention should be paid to the 3.3, 6.2, and 7.7~$\mu$m bands, among which the 7.7~$\mu$m feature will present the greatest challenge, as it can shift toward longer wavelengths within spectral classes B and C \citep{peeters2021spectroscopic}. Given Ariel's spectral range and resolution, PAH signatures beyond $\sim$7~$\mu$m are likely to remain below detection thresholds. Moreover, the visible slope accessible to the Ariel photometers, although potentially indicative of PAHs, degenerate with other opacity sources such as TiO, VO, and alkali metals. Synergistic observations with JWST, which provides higher spectral resolution and broader mid-IR coverage, and with high-resolution ground-based facilities, will be essential to disentangle these effects and confirm PAH detections \citep{changeat2025_synergy}.

\paragraph{Optimal Observing Modes}

Transmission spectroscopy across 0.5-5 $\mu$m is crucial: PAH-containing photochemical hazes can imprint (i) a super-Rayleigh optical slope shortward of 1 $\mu$m and (ii) the broad 3.3 $\mu$m C-stretch. In practice, PAH opacities are typically represented using empirical `astro-PAH' absorption cross-sections derived from \cite{li2001infrared} and \cite{draine2007infrared}, which capture the bulk optical behavior of PAH populations observed in astrophysical environments (e.g. \citealt{ercolano2022observations}). This treatment effectively describes the opacity of carbonaceous haze particles that may form from PAHs and related complex organics, analogous to the haze layers inferred in Titan's atmosphere. Ariel's AIRS-0 resolving power (R $\approx$ 100) is sufficient for these diffuse bands. Also, the intensity ratio of the 3.3 $\mu$m PAH band to the 3.4 $\mu$m aliphatic C-band feature can provide constraints on the chemical structures of the carriers and further hints about how H-rich the atmosphere is.

Emission spectra are decisive for hot (T$_\mathrm{eq}$ $\geq$ 1300K) targets where the 6.2-8 $\mu$m C-C features lie in AIRS-1; thermal contrast boosts the PAH signal and circumvents clouds that mute cooler atmospheres. At wavelengths longer than $\sim$ 5 $\mu$m, several PAH bands fall within the spectral coverage of JWST/MIRI and Ariel's AIRS-1 instrument. However, most exoplanet time-series observations with JWST use the MIRI low-resolution mode (R $\approx$ 40--160), where the noise floor is comparatively high, making robust identification of these features challenging. Ariel will probe a similar wavelength range with AIRS-1 but at lower spectral resolution (R $\approx$ 30--50), which may complicate the separation of PAH features from other hydrocarbons. Nevertheless, repeated observations of large planet samples with Ariel may help improve signal-to-noise and break degeneracies through population-level constraints. Overall, the detectability of PAHs in exoplanet atmospheres can be guided by the parameter-space investigation of \cite{arenales2025polycyclic}, which explores a range of PAH abundances (mass fractions of 10$^{-5}$, 10$^{-6}$, and 10$^{-7}$) using JWST observations of the 3.3 $\mu$m band and the optical slope. This study therefore provides a useful framework for evaluating the prospects of detecting PAH signatures with future Ariel observations.

Phase curves (Tier-3 cadence) on a handful of bright, short-period giants can reveal longitudinal PAH inhomogeneities and test photochemical pathways; optional but valuable.

\paragraph{Target selection criteria}

Detection limits (mass fraction of PAHs $\approx 10^{-7}$) are most favorable in hot Jupiter atmospheres under the following conditions. This value corresponds to the lowest PAH abundance explored in recent detectability simulations by \cite{arenales2025polycyclic} (where they explore PAH detectability for three different mass fractions, $10^{-5}$, $10^{-6}$, and $10^{-7}$), which were adopted to bracket plausible atmospheric scenarios motivated by PAH abundances inferred in the interstellar medium \citep{Tielens2008}.

\begin{itemize}
\item Equilibrium temperature 1100--1300 K: this is the sweet spot where thermally produced PAHs peak, while water still provides a relatively clean continuum.
\item High C/O ratio ($\geq 0.8$) and supersolar metallicity: although current C/O constraints remain model-dependent, early JWST spectra are already narrowing the allowed range.
\item Steep optical slopes in HST/JWST spectra: these indicate the presence of photochemical hazes that may contain PAHs.
\item Cooler CH$_4$-rich giants ($T_\mathrm{eq} < 900$ K) may also host PAHs. However, thermochemical models are pessimistic, and cloud opacity may bury their signatures. In such cases, simultaneous high-resolution ground-based spectroscopy would be required. Ariel should therefore prioritize the 1000--1300 K temperature window.
\end{itemize}

However, it is important to note that current thermochemical models remain incomplete with respect to PAH formation pathways. In particular, they often rely mainly on the hydrogen-abstraction/acetylene-addition (HACA) mechanism, which tends to underpredict PAH abundances at lower temperatures. Additional radical recombination processes, which become more efficient at cooler temperatures or in the presence of strong radiation fields, are not yet fully incorporated (see \citealt{kaiser2021aromatic} for a detailed review).\\
As a concrete example, WASP-6 b could be a good start with eclipse and transit observations. It has two critical parameters that are necessary for PAH observations: (i) T$_\mathrm{eq}$ = 1184 $\pm$ 16 \citep{tregloan2015transits} and {ii)} optical slope. The simulations show that PAHs are detectable down to 10$^{-3}$ $\times$ ISM abundance with one JWST transit \citep{grubel2025detectability}. It also is on the Mission Candidate Sample. New discoveries of TESS that meet the same criteria can enter Tier 1 reconnaissance and be upgraded as JWST refines their C/O ratios.


\subsection{Evolution of exoplanet atmospheres beyond the main-sequence}\label{sec:evolution}

\subsubsection{Context}
The large population of planets around main-sequence stars is already providing key insights about the planet formation process and how planetary systems are built to finally reach their present architecture. By contrast, very little is known about the evolution of these systems once the star ages towards the Red Giant Branch (RGB). Indeed, the life and properties of a planet and its atmosphere are closely linked to the life and properties of its host star \citep[e.g.][Sect. \ref{sec:young}]{Louca2025}. The relatively recent detection of transiting extrasolar planets beyond the main sequence (e.g., \citealt{lillo-box14,grunblatt24}) thus represents an opportunity to study the coupling between stellar and planetary evolution \citep{grunblatt23}. Among the more than 5000 planets identified around other stars, only 200 have been detected around evolved stars ascending RGB. And this relatively small population is dominated by long-period planets (detected by radial velocity) orbiting beyond 0.5 AU, with a paucity of planets in the close-in regions, pointing to the engulfment of close-in planets once the star evolves off the main sequence. However, a small population of $\sim$20 daring transiting exoplanets stands up to their giant parents and has been able to elude their fate. This population started with the detection of Kepler-91~b \citep{lillo-box14}. The door is then open to study the evolution of planetary systems as their host star ages to understand our own future. The launch of Ariel is a key opportunity to perform this study in a systematic way, exploring the atmospheres of planets around stars at different evolutionary stages.   

\subsubsection{Strategy for observations}

We propose to observe the panchromatic (vis-nir) transmission spectrum (low- and medium resolution) of five close-in planets around evolved stars at different evolutionary stages: from the sub-giant to the RGB phases. Since this niche has never been explored before, the goal is to investigate the species in the atmospheres of close-in planets around evolved stars and compare their abundances and interior structures (P-T profiles) with their counterparts in the main-sequence phase. 

\begin{figure}
	\centering
	\includegraphics[width=\linewidth]{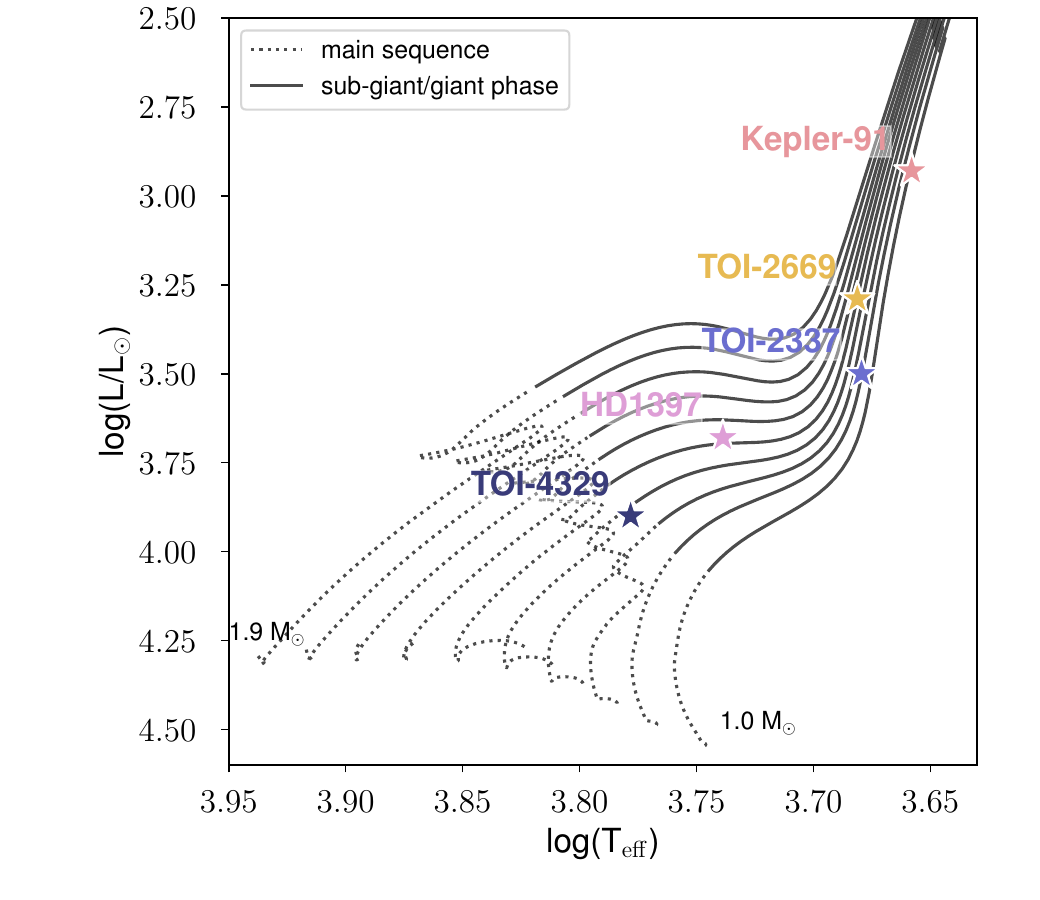}
	\caption{Hertzprung-Russell diagram zoomed into the post main-sequence parameter space. Evolutionary tracks from 1 to 1.9 M$_{\odot}$ are shown as lines, with the dotted regime corresponding to the sub-giant phase and the solid part to the red giant branch stage. The five targets marked, covering different stages of the evolutionary path, are among the most suitable close-in giant planets for atmospheric characterization with the Ariel mission.}
	\label{fig:TEMPO_targets}
\end{figure}

Atmospheric composition is a powerful proxy of the formation and migration history of planetary systems \citep[e.g.][Sect. \ref{sec:population-case1}]{welbanks19}. Differences in the chemical abundances (Na, K, H$_2$O) between planets around evolved stars and their main-sequence counterparts can reveal how evolution affects the properties of planetary atmospheres (e.g., \citealt{grunblatt23,kozakis20}). Additionally, atmospheric profiles change critically with stellar insolation \citep{Miguel2014, baxter20} so we should also see a transition in the P-T profiles of these planets. The strengths of these features also test theories of planet inflation and atmospheric mixing \citep{komacek19,baxter20}. According to the current population of known planets around evolved stars, the following planets would be very good targets for this study (in increasing evolutionary stage): TOI-4329~b, HD~1397~b, TOI-2337~b, TOI-2669~b, and Kepler-91~b (see Fig.~\ref{fig:TEMPO_targets}). They are already included in the list of potential Ariel targets defined in \cite{Edwards2022}.

\subsection{How does UV radiation from stellar flares affect the chemical composition of exoplanet atmospheres?}\label{sec:flares}

\subsubsection{Context}
The impact of stellar flares on exoplanets remains poorly understood. These energetic events are especially frequent in active stars, with M-type stars among the most active. Yet with no rigorous observations to back up theoretical studies, the influence of such high activity levels on planetary atmospheres is still unclear. One of the most anticipated effects is on atmospheric chemistry, particularly through UV-driven photochemistry, and on changes in the temperature profile, as suggested by several theoretical studies for small and large planets \citep{Segura2010, Venot2016, Chen2021, Konings2022, Louca2023, Ridgway2023, Nicholls2023}. In these studies, the flare spectra are not based on solar flares but on observations or reconstructions of flares from active stars, particularly M dwarfs \citep{France2016, 2018ApJ...867...71L}. Although stellar flares typically last only a few hours, the resulting photochemical perturbations after the flare, and the cumulative effect of continuous small flares, can persist longer and may lead to a modified chemical steady state of the observable region of the atmosphere.
While UV radiation is the primary driver of flare-induced photochemistry, X-ray ionization during flaring events can further enhance chemical complexity by producing secondary electrons that dissociate stable molecules and enable otherwise inaccessible ion-molecule reaction pathways. However, given the incomplete observational and theoretical constraints on X-ray-driven processes, their quantitative impact remains uncertain.
In particular, it has been shown that the steady-state abundances of species susceptible to photodissociation differ under the additional intense UV irradiation contribution from flare activity. Importantly, even species not directly sensitive to UV radiation can be affected due to the cascade of the chemical reactions triggered by flaring events, and the sudden availability of additional reactive species. Modeling studies have shown that it should be possible to observe the cumulative effects of flare-driven chemical changes using JWST-NIRSpec for warm Jupiter atmospheres \citep{Nicholls2023}. 

Besides the direct effect of a higher UV radiation on photochemically active species, 
flares may also influence atmospheric escape processes, particularly during strong or repeated events that will also affect the chemistry in longer timescales \citep{Louca2023}. Furthermore, \cite{Pass2025} have framed the potential for flares to enhance cumulative photoevaporative losses in the context of the \textit{cosmic shoreline} \citep{Zahnle2017}. The cosmic shoreline demarcates the boundary between airless planets and those that possess an atmosphere, providing critical guidance for telescope surveys such as JWST DDT program on Rocky Worlds\footnote{https://rockyworlds.stsci.edu/}.

\subsubsection{Strategy for observations}

Directly observing the effects of stellar flares on exoplanet atmospheres is a complex challenge and requires repeated observations of the same system to capture atmospheric variability associated with stellar activity. One potential strategy is to carry out repeated transmission spectroscopy of exoplanets orbiting active stars, aiming to detect atmospheric variability over time. Ideally, this would involve obtaining measurements of a planet during three successive phases: (1) when the star is quiet, to establish a baseline scenario; (2) shortly after a flare or during a period of sustained flare activity, to measure photochemical or thermal responses; and (3) through follow-up observations during a quiescent period, to assess whether the atmosphere returns to its pre-flare state. Implementing a strategy to capture flares in real-time would require a rapid alert system and the flexibility for Ariel to interrupt its planned schedule-capabilities that may not currently exist. However, since modeling studies have established that successive flares can cumulatively perturb steady-state compositions, comparisons between quiet- and active-periods are both theoretically and observationally justified.

Furthermore, the atmospheric response to a flare may not be immediate. Chemical and thermal disequilibrium can develop over hours to days, and in some cases, the atmosphere may settle into a new steady state rather than reverting to the original one \citep[e.g.][]{Konings2022, Louca2023}. As such, Ariel is best suited for post-flare monitoring, capturing long-term photochemical effects through its multi-epoch survey strategy.

Recent modeling studies support the observability of such effects: for instance, a moderate flare ($\sim$2$\times$10$^{33}$ erg) can alter transmission spectra by 100-300 ppm in CH$_4$ and NH$_3$ bands, and by up to 350 ppm in C$_2$H$_2$ near 14 $\mu$m \citep{Konings2022}. These variations are within Ariel's detection capabilities, especially for planets with deep transits and repeated visits.

We therefore recommend that Ariel prioritize a small number of targets orbiting flare-active M dwarfs, such as: AU Mic b (short-period planet around a very active M1 dwarf with frequent UV superflares; ideal for cumulative flare effects and escape processes), GJ 1214 b (warm sub-Neptune orbiting an M4 star with known UV activity; a strong candidate for observing CH$_4$ and NH$_3$ photochemical evolution, although clouds and hazes may mute spectral features in transmission spectra), TOI-1452 b (a transiting super-Earth around an M4 star with confirmed optical flares detected by TESS; suitable for testing high-energy irradiation models).

Long-term monitoring of such systems may allow Ariel to constrain the role of flares in atmospheric evolution, explore time-dependent photochemistry, and test predictions on the resilience or instability of planetary atmospheres under variable stellar activity. This approach is complementary to Ariel's primary goals and will enhance our understanding of the habitability and chemical diversity of planets around low-mass stars.

\subsection{Observing the signature of energetic particle-driven photochemistry in exoplanet atmospheres}\label{sec:energetic-particles}

\subsubsection{Context}
Observations of SO$_2$ detected with JWST \citep[e.g.][]{Tsai2023b, Dyrek2024} demonstrate that photochemistry plays an important role in exoplanet atmospheres (see Sect.\ref{sec:disequilibrium}). In addition to high-energy stellar radiation, energetic particles likely drive chemistry in exoplanet atmospheres. Energetic particles are relativistic particles accelerated by the exoplanet's host star (known as stellar energetic particles), via flares and coronal mass ejections, and cosmic rays from the Galaxy. Energetic particles are important for exoplanets because, for instance, chemical models have shown that energetic particles can drive the formation of prebiotic molecules in Earth-like \citep{airapetian-2016} and Jupiter-like atmospheres \citep{barth-2021,rodgers-lee-2023}. Investigating prebiotic molecule formation is important for our understanding of the origin of life-as-we-know-it on Earth and the potential for it to develop on other exoplanets.

While it seems unlikely for life to develop on gas giant atmospheres, we can use them to learn about atmospheric chemistry \citep{rimmer-2023}. Energetic particles are by no means the only mechanism by which prebiotic molecules can be formed; UV radiation could also lead to the formation of some prebiotic molecules \citep{Harman2013,Green2021}. While the air showers from high-energy cosmic rays ($>$GeV energies) penetrate deep in the atmosphere, lower energy cosmic rays will lose their energy higher in the atmosphere. The same is also true for stellar energetic particles since their spectra peak at low energies. In order to isolate the signature of low-energy ($<$GeV energies) energetic particles in hydrogen-dominated gas giant atmospheres, chemical signatures that are thought to be most likely caused by energetic particles have been identified using chemical models as \ce{H3+} and its products with abundant background molecules such as water, leading to \ce{H3O+} and \ce{NH3}, leading to \ce{NH4+} \citep{helling-2019, barth-2021}. These molecules have absorption features in the near-infrared. Ariel, operating in the infrared, is an excellent instrument to search for energetic particle chemistry on exoplanets because Ariel will observe a large number of exoplanets and will provide constraints to stellar properties that are needed to robustly connect planet-star interactions required for energetic particle chemistry. While X-rays may drive similar chemistry, this chemistry should dominate higher in the atmosphere where the effects are less likely to be observed. X-ray photochemistry is included in \citet{barth-2021} but stellar energetic particles are still identified as the main drivers of \ce{H3+}, \ce{H3O+} and \ce{NH4+} formation.

\begin{figure*}
\centering
\includegraphics[width=0.9\linewidth]{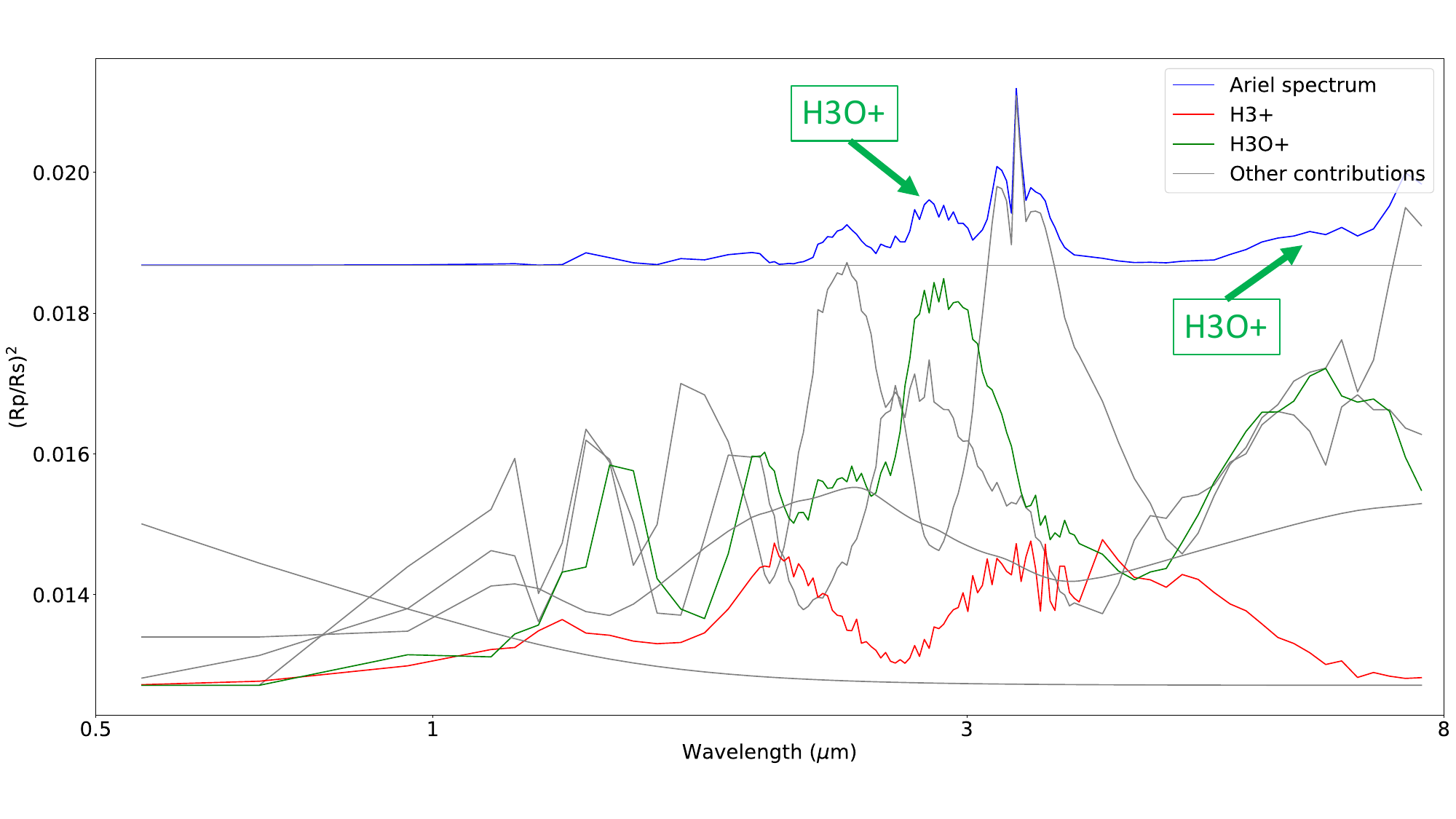}
\includegraphics[width=0.9\linewidth]{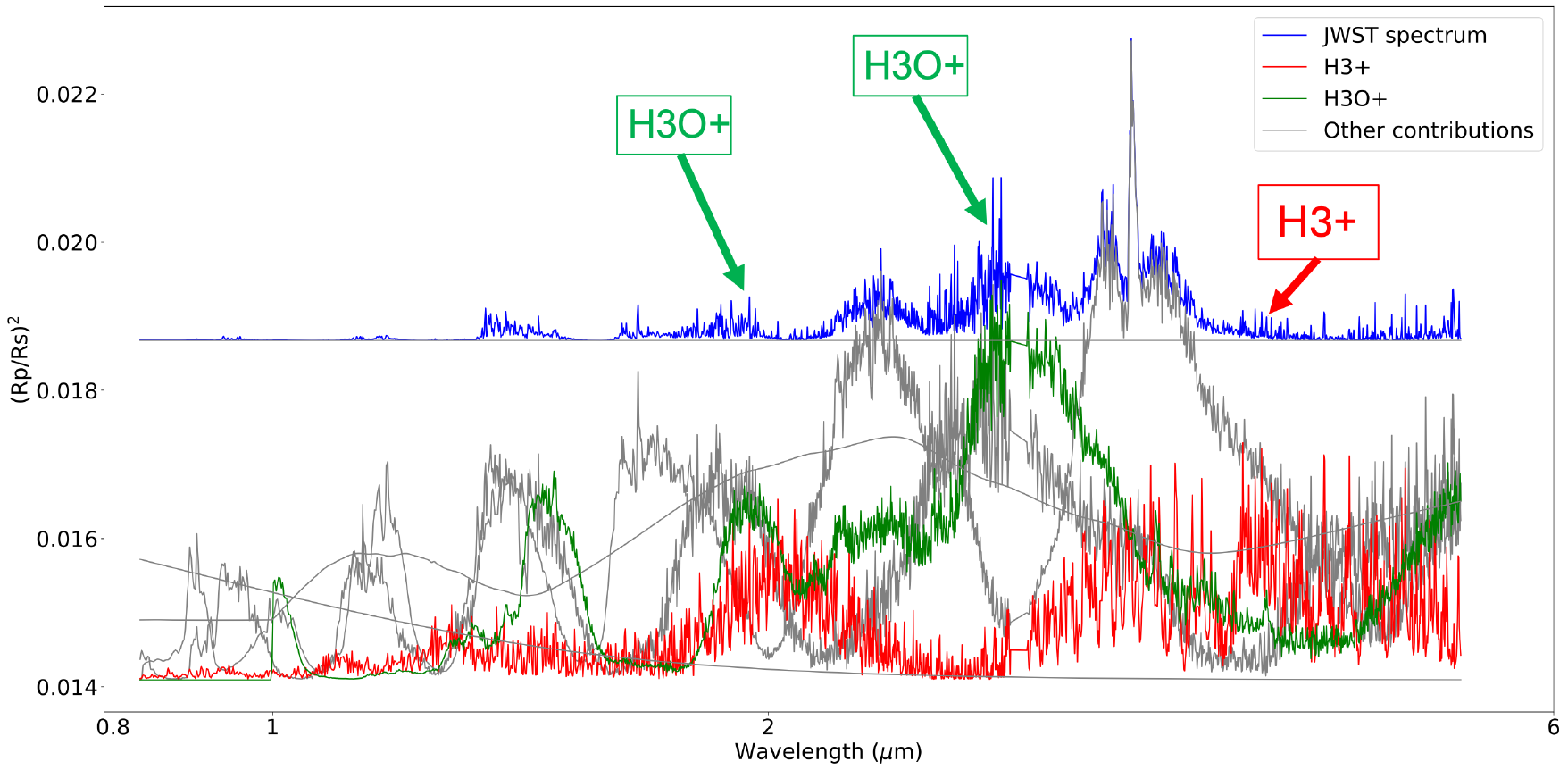}
\caption{\label{fig:bourgalais} Illustrative simulated transmission spectra of a GJ~1214b-like atmosphere assuming representative \ce{H3O+} abundances adopted by \citet{bourgalais-2020}. The simulations are intended to demonstrate the detectability of ionospheric tracers with Ariel and JWST rather than to provide a self-consistent atmospheric model for GJ~1214b.}
\end{figure*}

\subsubsection{Strategy for observations}

Illustrative synthetic Ariel transmission spectra presented by \citet{bourgalais-2020}, based on laboratory evidence for ion production together with representative atmospheric ion abundances inspired by ion-chemistry models, indicate that the broad absorption feature at 2.8,$\mu$m due to \ce{H3O+} would be detectable if present at these abundance levels. While these simulations are not intended as self-consistent models of a specific planet such as GJ~1214b, they demonstrate Ariel's sensitivity to ionospheric tracers and motivate further modelling of energetic particle chemistry in exoplanet atmospheres.
The ion \ce{H3O+} provides a probe of the physics and chemistry of exoplanet ionospheres by tracing both the energetic particle ionisation rate and the water content in the upper atmosphere. If two of the following are known -- (a) water vapour concentration, (b) \ce{H3O+} concentration, or (c) cosmic ray ionisation rate -- then the third can be inferred. Thus, if \ce{H2O} and \ce{H3O+} concentrations are measured, the cosmic ray ionisation rate can be predicted; conversely, if the ionisation rate is constrained from stellar magnetic maps and wind models, \ce{H3O+} observations allow the water content of the ionosphere to be determined, providing a measure of the above-cloud C/O ratio in giant exoplanets. 

While \ce{H3+} has been detected in 
the mid-infrared on Jupiter \citep{stallard-2001}, Saturn \citep{geballe-1993} and Neptune \citep{trafton-1993}, it has not been detected on hot Jupiters thus far. \citet{bourgalais-2020} indicate that warm Neptune atmospheres may be more favourable to observe \ce{H3+}, rather than hot Jupiters but that \ce{H3+} still shows much weaker features than \ce{H3O+} that would be largely masked by clouds. They suggest that higher \ce{H3+} abundances than they assumed would be needed, or a larger number of orbits than is needed for \ce{H3O+}, to detect \ce{H3+} with Ariel.

The most favourable targets are warm exoplanets (T$_{\mathrm{eq}}<1200$ K). Short-period planets orbiting young or active stars enable studies of stellar energetic particles, while long-period planets are sensitive to Galactic cosmic rays (possibly with emission spectroscopy). Promising targets also include systems where the stellar large-scale magnetic field strength is known and stellar wind models are available, since these constrain energetic particle transport. The resulting ionisation rate is a key input for models of particle-driven chemistry. From the Mission Candidate Sample, the best exoplanet targets meeting these criteria (T$_{\mathrm{eq}}<1200$ K, magnetic maps, and stellar wind models) are HN~Peg~b, GJ~436~b, AU~Mic~b, HD~130322~b, HD~63433~b,~c, and HD~189733~b. Further chemical modelling, including synthetic Ariel spectra of these planets, will be performed and would be valuable to determine how many orbits are required to observe energetic particle-driven photochemistry.


\subsection{Get to know the Giant planets around M dwarfs}\label{sec:giants-Mstars}

\subsubsection{Context}
Studying giant planets around M stars with Ariel offers a unique and compelling science case. In the classical core accretion scenario, giant planet formation requires substantial amounts of gas and solids in the protoplanetary disk. Since disk mass is generally thought to scale with stellar mass, it has long been assumed that giant planets are unlikely to form around low-mass stars like M dwarfs \citep{Miguel2020}. However, recent discoveries from TESS and other surveys have revealed a growing population of giant planets orbiting M stars \citep{Bryant2024}, with preliminary estimates of planetary occurrence rate of 0.194$\pm$0.072 
\citep{Bryant2023}. Additionally, most stars found to host giant planets are metal-rich \citep{Gan2025}. Verifying whether the planets themselves are similarly enriched in metals would help test and refine the core accretion theory as the dominant formation pathway for these worlds. Close-in giant planets around M dwarfs are also preferentially found in S-type binary systems with widely-separated stellar companions \citep{Frensch_2026}, which points towards post-disk, Lidov-Kozai migration as potentially the dominant mechanism of close-in gas giant formation around M dwarfs \citep{Weisserman_2025}. Disk-driven versus post-disk driven migration pathways are thought to produce differing chemical compositions in the atmospheres of these planets \citep[e.g.][]{Molliere2020,Feinstein_2025}, which will be traced with Ariel. Taken together, these findings not only challenge current models of planet formation but also offer an important opportunity to deepen our understanding of the processes involved \citep{Stefansson2023, Gan2023}.

M stars are often active, producing frequent flares and intense stellar radiation. Repeated observations of giant planets around these stars can provide additional key insights into how stellar activity affects planetary atmospheres (see Sect. \ref{sec:flares}). On top of that, the relatively large size of giant planets compared to their small host stars makes them ideal for transit spectroscopy, maximizing the atmospheric signal Ariel can detect. Altogether, these factors make giant planets around M stars excellent targets for the mission.

\subsubsection{Strategy for observations}
The main deliverable to be obtained from the observations are the atmospheric abundances. In particular, the metallicity in their atmospheres would inform on potential deviations from the stellar metallicity and would also help to determine the planet formation environment where the planets were born. Given the relative difference between these large planets and their small host stars, transit spectroscopy is the most recommended observational technique. 

Several of these planets are scheduled to be observed with JWST, and for those, repeated observations with Ariel would help to determine potential changes in their atmospheric chemistry due to the fact that they orbit potentially active stars. These planets are: TOI-3984~b, TOI-3757~b, HATS-6~b, HATS-75~b, TOI-5293~b, TOI-3714~b, TOI-5205~b \citep{kanodia2024}. Among these, HATS-6~b and TOI-3714~b are also included in the Mission Candidate Sample, offering a unique opportunity to conduct this comparative exercise between JWST and Ariel, and to investigate potential temporal variations in their atmospheric chemistry. 

Other planets have been detected with TESS and have no atmospheric detections to date. Here is a list of additional 26 giant planets with mass determination in addition to the transit detections with TESS \citep{Gan2025, kanodia2024}: GJ~1148~b, GJ~1148~c, GJ~179~b, GJ~3512~b, GJ~649~b, GJ~849~b, GJ~849~c, GJ~876~b, GJ~876~c, HIP~79431~b, Kepler-45~b, NGTS-1~b, TOI-1899~b, TOI-3629~b, TOI-3714~b, TOI-3757~b, TOI-4201~b, TOI-519~b, TOI-5205~b, TOI-530~b, K2-419~A~b, TOI-6034~b. Most of these planets are already included in the MCS, representing a valuable sample for future atmospheric characterization and enabling Ariel to significantly expand its statistical coverage of TESS-discovered temperate and warm gas giants.

\subsection{Probing Giant Planet Formation Through Atmospheric Characterisation in Young Systems }\label{sec:young}

\subsubsection{Context}
To date, most atmospheric characterizations have focused on mature exoplanets orbiting main-sequence stars \citep[e.g.][]{Sing2016, Welbanks2019, mansfield2021}. However, these planets have undergone billions of years of atmospheric evolution, including hydrodynamic escape, chemical processing, and potential modification by stellar activity. Consequently, it remains unclear to what extent the observed compositions reflect their formation environments versus subsequent atmospheric reshaping \citep[e.g.][]{Fortney2013, Louca2023b}. Interpreting formation pathways from such atmospheres is therefore highly degenerate and strongly model-dependent.

In contrast, giant planets in very young systems (ages $<$ 100 Myr) offer a rare chance to observe atmospheres that are still close to their primordial state. These `adolescent' planets have not yet experienced the full effects of long-term atmospheric loss and stellar irradiation, making them excellent probes of initial bulk composition and formation conditions. By comparing the chemical fingerprints, such as metallicities and elemental ratios of young and evolved giants, we can begin to trace how planetary atmospheres change over time and disentangle formation signatures from evolutionary effects.

\subsubsection{Strategy for observations}
The Ariel mission, with its broad wavelength coverage and high sensitivity to key molecular features, is uniquely positioned to lead this investigation. Since these planets would be less affected by evolution, the C/O ratio derived from the Ariel transit measurements of \ce{CO}, \ce{CH4}, and \ce{H2O} is our best chance to learn about the formation location of young transiting planets with respect to the primordial protoplanetary disk ice lines.

Furthermore, Ariel's ability to observe a statistically significant sample of exoplanet atmospheres, including a subset of young, transiting gas giants, will enable systematic studies of atmospheric composition across different evolutionary stages. This will provide critical insights into the physical processes driving planet formation, atmospheric retention, and chemical evolution.

\begin{figure*}
\centering
\includegraphics[width=0.8\linewidth]{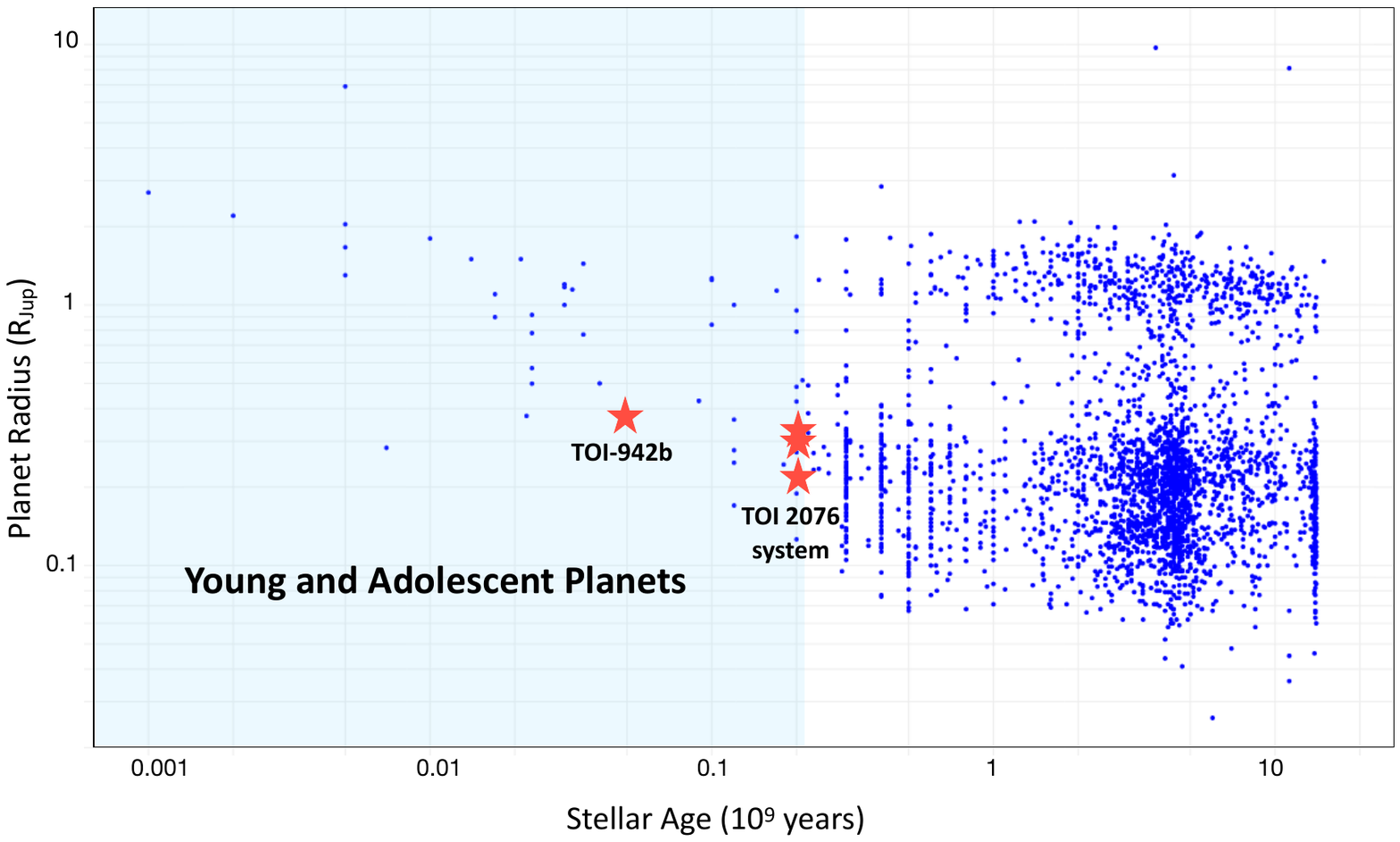}
\caption{Stellar age versus planet radius for all planets in the Mission Candidate Sample. Planets younger than 200 Myr are highlighted in blue. The four planets marked with a star represent promising young targets.
}\label{fig:youngplanets}
\end{figure*}

As shown in Figure~\ref{fig:youngplanets}, there are close to 40 planets currently on the Mission Candidate Sample with an estimated age of $<$ 200 million years. Among them, we can mention two interesting systems as examples. First, TOI-942~b, which is about 50 million years old. Second, the planets of the TOI-2076 system, to be observed by JWST in the KRONOS program (PI: Feinstein). Repeated observations would be important to test the evolution of their atmospheres over time, especially since these young planets orbit active stars (Sect.~\ref{sec:flares}). Except for TOI-2076~b, the planets from this program are currently included in the MCS, making this system a particularly good example of a synergistic study between Ariel and JWST.

\subsection{Tracing Planet Formation via Atmospheric Composition}\label{sec:planet_formation}

\subsubsection{Context}

One of the central goals of the Ariel mission is to understand how planets form and evolve by analyzing the composition of their atmospheres. Giant planets, in particular, provide a powerful window into formation processes, as their extended atmospheres may retain signatures of their accretion histories and migration pathways \citep{Oberg2011, Turrini2021}. However, as highlighted in Sect.~\ref{sec:young} and throughout this manuscript (Sects.~\ref{sec:evolution}, \ref{sec:flares}, \ref{sec:energetic-particles}), linking atmospheric composition directly to formation scenarios remains challenging due to the complex interplay of disk chemistry, accretion, migration, planetary and stellar evolution, and interior-atmosphere coupling \citep{molliere2022, Venot2016, Chen2021, Konings2022, Nicholls2023, Miguel2022, Louca2023, Louca2023b, Louca2025}.

In collaboration with the Ariel Planet Formation Working Group, we explored how different formation histories translate into observable atmospheric signatures, adopting the simplifying assumption of fully mixed interiors and no subsequent evolution. The study simulated the formation, migration and subsequent chemistry in the atmosphere of 48 giant planets under a wide range of conditions. While the full study is presented in another accompanying publication in this special edition \citep{Miguel2026}, we discuss here our main results to show the implications for Ariel. We used two potential outcomes for the envelope enrichment in these planets: solid-enriched planets are those in which the atmospheric metallicity is primarily determined by the dissolution of accreted planetesimals into the envelope, while gas-dominated planets acquire most of their atmospheric composition from the chemical makeup of the gas accreted from the protoplanetary disk, resulting in systematically different elemental ratios and molecular abundances in their atmospheres. In addition to these differences, we also considered, varying disk compositions (inherited vs. reset), ionization levels (high and low), and formation distances (5, 50, and 130 AU) \citep{Turrini2021, Pacetti2022}, with subsequent migration to either 0.1 or 0.4 AU. Atmospheric spectra were  modeled to evaluate their potential observability with Ariel. 

The results show that variations in the primordial disk chemistry or ionization state produce only very small differences in the transmission spectra, making these scenarios difficult to distinguish observationally since the differences are smaller than the error bars of Ariel (or even JWST) measurements. Instead, the dominant observable differences arise from the mechanism of envelope enrichment and from the planet's final orbital location, which controls atmospheric temperature and chemistry. For example, for planets migrating to 0.1 AU ($\sim$900 K), differences in formation location can produce detectable CO$_2$ features near 4.3 $\mu$m when planets form farther from the star at 130 AU. In contrast, cooler planets that migrate to 0.4 AU ($\sim$450 K) show spectra dominated by CH$_4$ and NH$_3$ features, largely erasing the CO$_2$ formation tracer even if they originally formed at 130 AU. The enrichment mechanism, however, leaves stronger spectral signatures: solid-enriched atmospheres display enhanced H$_2$O and NH$_3$ absorption bands, whereas gas-dominated atmospheres are characterized by stronger CH$_4$ features. These differences lead to spectral variations of order $\sim$100--300 ppm in transit depth across major molecular bands, making them potentially distinguishable with current facilities such as JWST and, in some cases, with Ariel through repeated observations. Ariel, with its broad wavelength coverage and sensitivity to key molecular species, is ideally positioned to test this prediction across a statistically significant sample of transiting giant planets.

\subsubsection{Strategy for observations}

Given these insights, Ariel is ideally positioned to test formation models by targeting transiting giant planets whose atmospheres may still preserve the fingerprints of their accretion histories, which, as detailed in Sect.~\ref{sec:young}, are more likely to be observed in young and adolescent planets. The mission's broad wavelength coverage and sensitivity to key molecular features such as \ce{H2O}, \ce{NH3}, \ce{CH4}, and \ce{CO2} make it possible to distinguish between solid-enriched and gas-dominated scenarios, especially in hot planets where spectral features are stronger and atmospheric scale heights are larger.

A productive observational strategy for Ariel would involve prioritizing Jupiter-sized exoplanets transiting relatively close to their stars, particularly those with equilibrium temperatures around or above 900 K. These planets offer the strongest contrast between formation scenarios, as higher temperatures amplify the spectral differences caused by varying chemical abundances. Cooler planets, by contrast, tend to be more compact and exhibit muted spectral features, making it harder to discern subtle differences.

Young gas giants, whose atmospheric compositions are less affected by long-term processes such as hydrodynamic escape or chemical reprocessing, are especially valuable targets (see Sect.~\ref{sec:young}). Comparing such planets to older counterparts observed with Ariel will allow for a systematic investigation of how atmospheric metallicity evolves over time and whether any observed enrichment can be attributed to formation or to subsequent atmospheric evolution. Particular attention should be given to planets whose bulk compositions or masses are already constrained through other methods, as these provide critical anchor points for testing and refining theoretical models.

Ultimately, by focusing on planets spanning a range of formation environments and accretion histories, Ariel can deliver a statistically meaningful test of how solid accretion shapes the atmospheres of gas giants and, more broadly, advance our understanding of the fundamental processes that govern planet formation.

\subsection{How do hot Neptunes survive in the hot-Neptune Desert?}\label{sec:hot-Neptunes}

\subsubsection{Context}

There is a scarcity of Neptune-sized planets in very short orbits ($\lesssim$ 5 days), a region known as the ``Neptune Desert'' \citep{Lundkvist2016,Mazeh2016}. Atmospheric loss has been proposed as one of the processes responsible for creating this desert. Nevertheless, a handful of Neptune-sized planets have been discovered within this region, particularly by TESS \citep{Jenkins2020,Torres2024}. These rare survivors provide valuable insights into the origin of the Neptune Desert.

Despite its importance, the atmospheric composition of these planets remains largely unconstrained. One ultra-hot Neptune of particular interest, LTT~9779~b, shows a muted spectrum consistent with high metallicity and/or high-altitude clouds \citep{Radica2024}. Silicate clouds have been proposed both to explain the suppression of atmospheric loss and the observed high albedo of LTT~9779~b \citep{Hoyer2023}. The gas-phase absorption of SiO within Ariel's spectral range could provide an independent confirmation of silicate species. Furthermore, no sulfur species have yet been detected in the limited atmospheric data available for hot Neptunes, even though sulfur is considered a promising tracer of metallicity and photochemistry (Sect.~\ref{sec:sulfur}; \citealt{Tsai2023b,Crossfield2023}). Constraining the atmospheric composition of these planets can therefore help test the hypothesis that they share a common Jovian origin.

Here, we aim to address two key questions:
\begin{itemize}
    \item What are the distinctive atmospheric features of hot Neptunes in the Neptune Desert?
    \item How do hot Neptunes evolve from the edge of the Neptune Desert (i.e. the Neptune Savanna; \citealt{CastroGonzalez2024}) toward its center, and do they follow the same mass-metallicity relationship as Solar System planets?
\end{itemize}

\subsubsection{Strategy for observations}
\begin{figure}
\centering
\includegraphics[width=\linewidth]{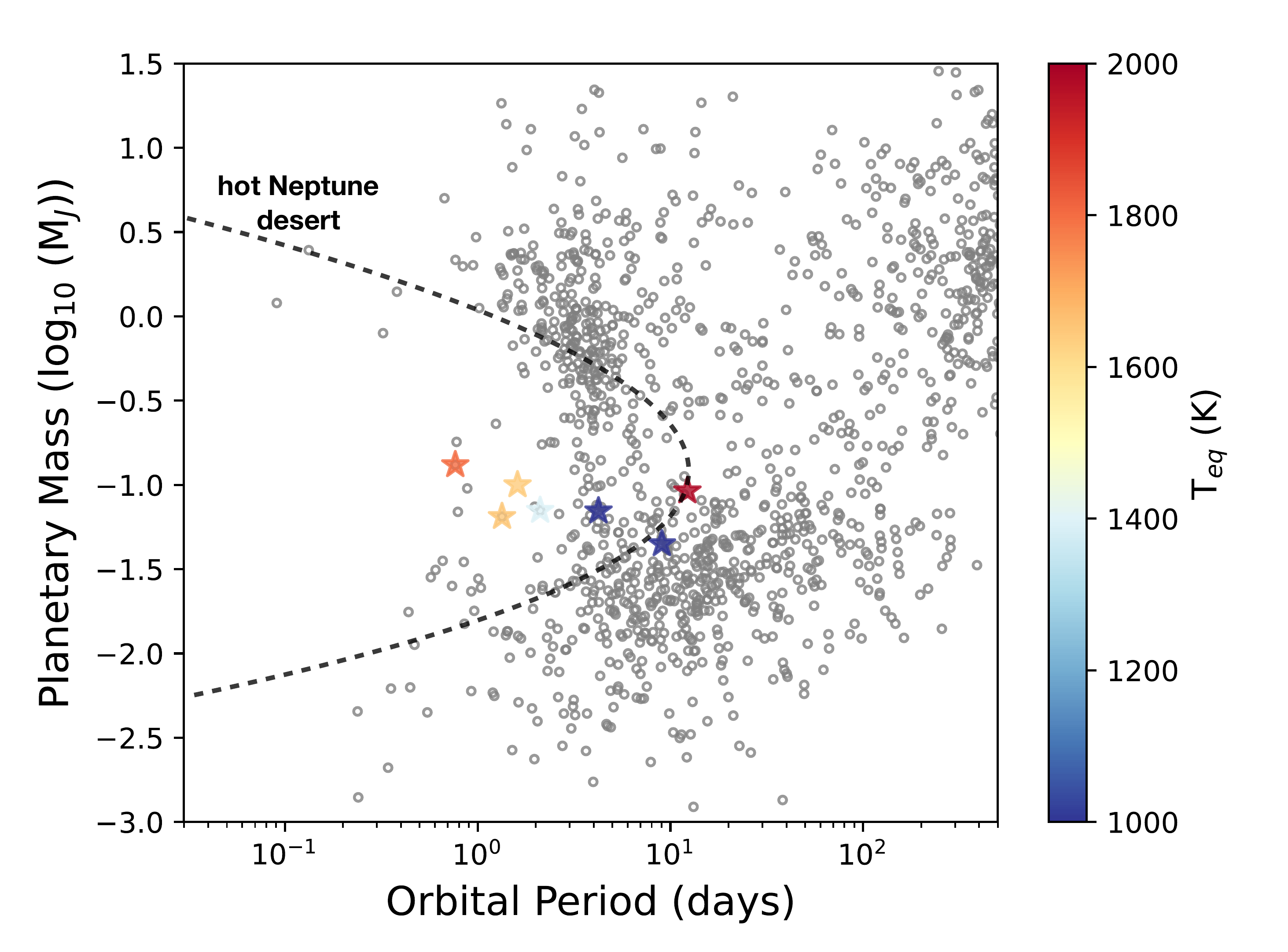}
\caption{Planetary masses vs orbital periods for exoplanets with mass constraints, taken from NASA archive (NASA~2024.09.17). Stars are the proposed targets spanning across the center to the end of the desert, color-coded with equilibrium temperature. The dashed line represents the boundaries of the Neptunian desert, following \protect\cite{Mazeh2016}.}
\label{fig:MP}
\end{figure}

Ariel's broad spectral range (0.5-7.8 $\mu$m) is essential for characterizing key molecular features. In particular, \ce{SO2} (7-8 $\mu$m), \ce{SiO} (4-4.3 $\mu$m), and \ce{CO2}  (4.2 $\mu$m) would be essential for constraining atmospheric metallicity and inferring their evolution paths. Given that existing data for LTT 9779 b suggest the presence of clouds, we propose both transmission and emission spectroscopy to probe different atmospheric regions and mitigate the impact of clouds on transmission spectra. We will focus on a short list of targets across the Neptune Desert and the Neptune Savanna -- a moderately populated region at orbital periods $\gtrsim 5$ days -- \citep{CastroGonzalez2024}, as shown in Figure~\ref{fig:MP}. The planets of interest for this study are TOI-849~b, LTT~9779~b, NGTS-4~b, TOI-132~b, WASP-166~b, and WASP-47~d. However, some of these planets (TOI-849~b, NGTS-4~b, and TOI-132~b) cannot be included in the Mission Candidate Sample because, although their host stars can be observed, too many visits would be required to achieve Tier 2 data quality.

This small survey aims to trace the transition from the edge to the center of the Neptune desert.

\section{Sub-Neptunes, super-Earths and rocky planets}
In this Section we focus the discussion to the science cases that involve small sub-Neptunes, super-Earths, and rocky planets. 

\subsection{Uncovering the bulk compositions of sub-Neptunes to test Radius Valley emergence mechanisms across the main sequence}\label{sec:sub-Neptunes-evolution}

\subsubsection{Context}
Around Sun-like stars and M dwarfs alike, the radii of small close-in planets form a bimodal distribution --known as the Radius Valley-- separating terrestrial super-Earths from the larger, low-density sub-Neptunes. The bulk compositions of individual sub-Neptunes cannot be gleaned in the absence of atmospheric measurements due to degeneracies with interior structure models when only planet mass, radius, and temperature measurements are available. Thus, a population-level demographic study with Ariel is needed to shed light on sub-Neptune compositions, and consequently on their origins (see also Small Planet WG Report 2026, in prep.).

The emergence of the Radius Valley has long been attributed to thermally driven atmospheric escape that erodes the primordial H$_2$/He atmospheres of sub-Neptunes, leaving behind bare rocky cores \citep[e.g.][]{Owen_2017,Gupta_2019}. The observed slope of the Radius Valley in radius-instellation space supports this hypothesis around FGK stars \citep{Fulton_2017,VanEylen_2018,Martinez_2019,Petigura_2022}, with planets' gas accretion histories setting the initial conditions for thermal escape \citep{Lee2022b,Nielsen2025}. 

However, the slope of the Radius Valley around early M dwarfs is markedly shallower \citep{Cloutier_2020,Gaidos_2024,Ho_2024}. This behaviour has been interpreted as evidence that planet formation may play a more prominent role around low-mass stars than around FGK stars, with atmospheric escape acting as a secondary process \citep{Burn_2024}. In this scenario, many sub-Neptunes would be water-rich rather than predominantly H$_2$/He dominated.

Additional observational evidence that has been interpreted as being consistent with a population of water-rich worlds around M dwarfs comes from detailed characterization of individual planetary systems \citep[e.g.][]{DiamondLowe_2022,Piaulet_2023}, planet population-level studies \citep[e.g.][]{Luque_2022,Cherubim_2023,Gillis2026}, and planet formation models that can reproduce the Radius Valley with water-rich formation and migration \citep{Venturini_2020,Venturini_2024, Burn_2021,Burn_2024}. We note that the effects of atmospheric escape are included in these models, but atmospheric escape is less impactful on the population of steam-dominated water worlds compared to H$_2$/He-dominated sub-Neptunes. 
Taken together, these studies motivate the hypothesis that close-in sub-Neptunes around FGK stars may preferentially retain H$_2$/He envelopes, whereas a larger fraction of sub-Neptunes around M dwarfs may instead be water-rich worlds with predominantly steam-dominated atmospheres. Ariel is uniquely positioned to test this hypothesis.
Nevertheless, we note that sub-Neptunes could also retain hydrogen-rich atmospheres, as suggested by several recent studies \citep[e.g.][]{Benneke2019_GJ3470b, Rogers2025, Owen_2017, Bean2021}, which highlights the importance of distinguishing between H$_2$-dominated gas dwarfs and steam-dominated water worlds. In addition, recent modeling of rocky super-Earths shows that photochemistry is required to explain the JWST inference of SO$_2$ in otherwise hydrogen-dominated atmospheres \citep{nicholls2025}. Since similar pathways have been invoked for WASP-39 b, this provides an additional motivation for Ariel to test photochemical effects in warm super-Earths and sub-Neptunes.

Ariel could test these scenarios by targeting a suite of sub-Neptunes around a variety of host star spectral types to distinguish high versus low mean molecular weight atmospheres. Uncovering the fraction of sub-Neptunes whose atmospheres are H$_2$/He-dominated versus steam-dominated across the main sequence has the potential to drastically improve our understanding of how super-Earths and sub-Neptunes form across the main sequence. 

\subsubsection{Strategy for observations}
We propose a program with Ariel to obtain Tier 2-level transmission spectra of a statistically significant sample of sub-Neptunes orbiting a variety of host star spectral types (i.e. from mid-M to G dwarfs) to distinguish H$_2$/He-dominated atmospheres from heavy volatile-rich atmospheres. Regarding target selection, we begin by querying the NASA Exoplanet Archive for planets with $R_\mathrm{p} \in [1.8,2.8]\, R_\oplus$ or $<1.8\, R_\oplus$ with mass and radius measurements that reveal a lower bulk density than a purely rocky body at the planet's mass. Our upper size limit of $2.8\, R_\oplus$ is informed by interior structure models of planets with a 50\% water mass fraction over a range of equilibrium temperatures $T_{\mathrm{eq}}$ in [400-1000] K \citep{Aguichine_2021}. We expect that sub-Neptunes smaller than $2.8\, R_\oplus$ can plausibly be explained by a thick water layer, whereas above $2.8\, R_\oplus$, a H$_2$/He envelope is needed to explain the planet's mass and radius. We also limit our targets to those with $T_{\mathrm{eq}}<900$ K to avoid targets that are largely susceptible to efficient thermally-driven atmospheric escape.

\begin{figure*}
\centering
\includegraphics[width=0.49\textwidth]{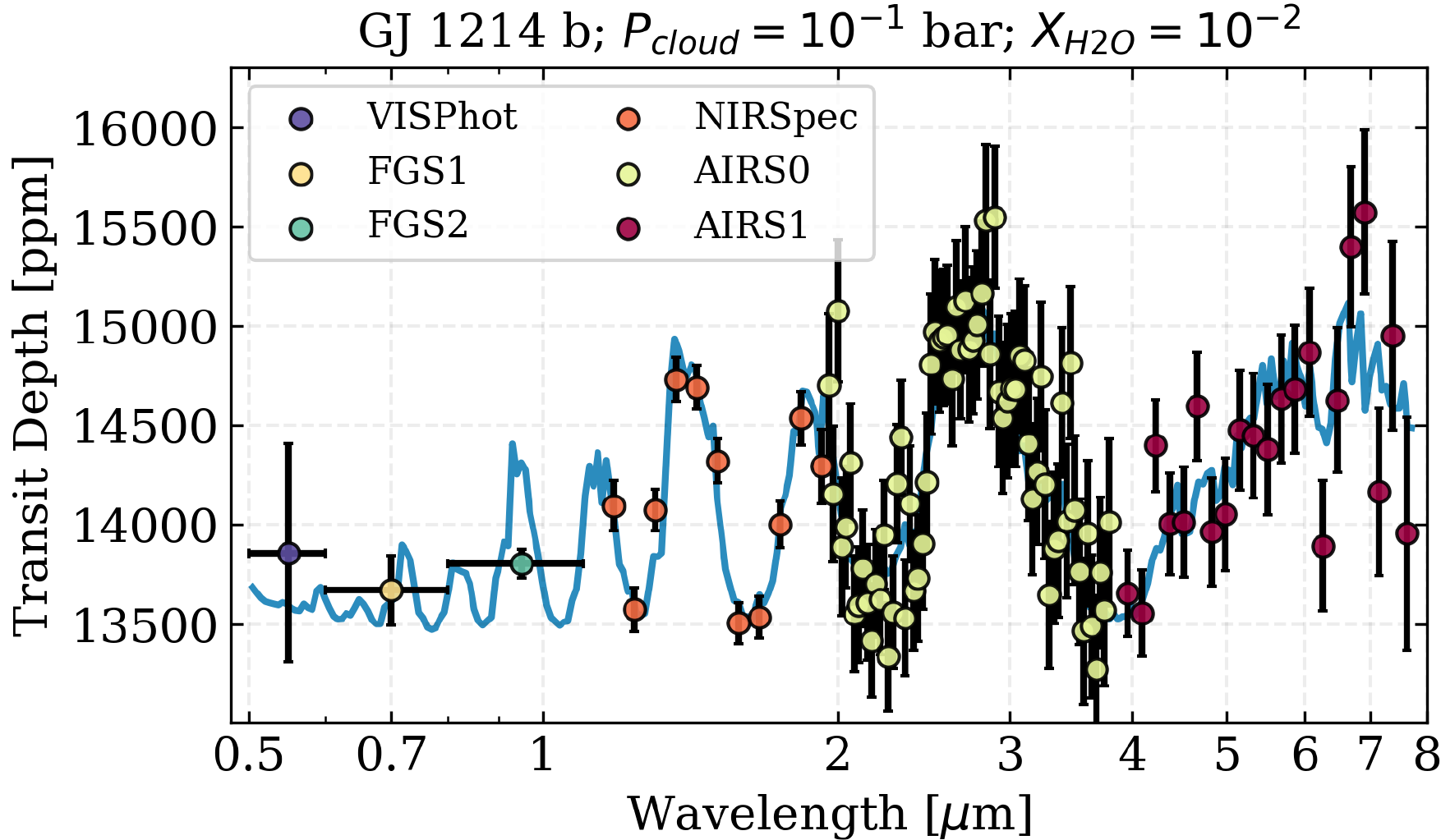}
\includegraphics[width=0.49\textwidth]{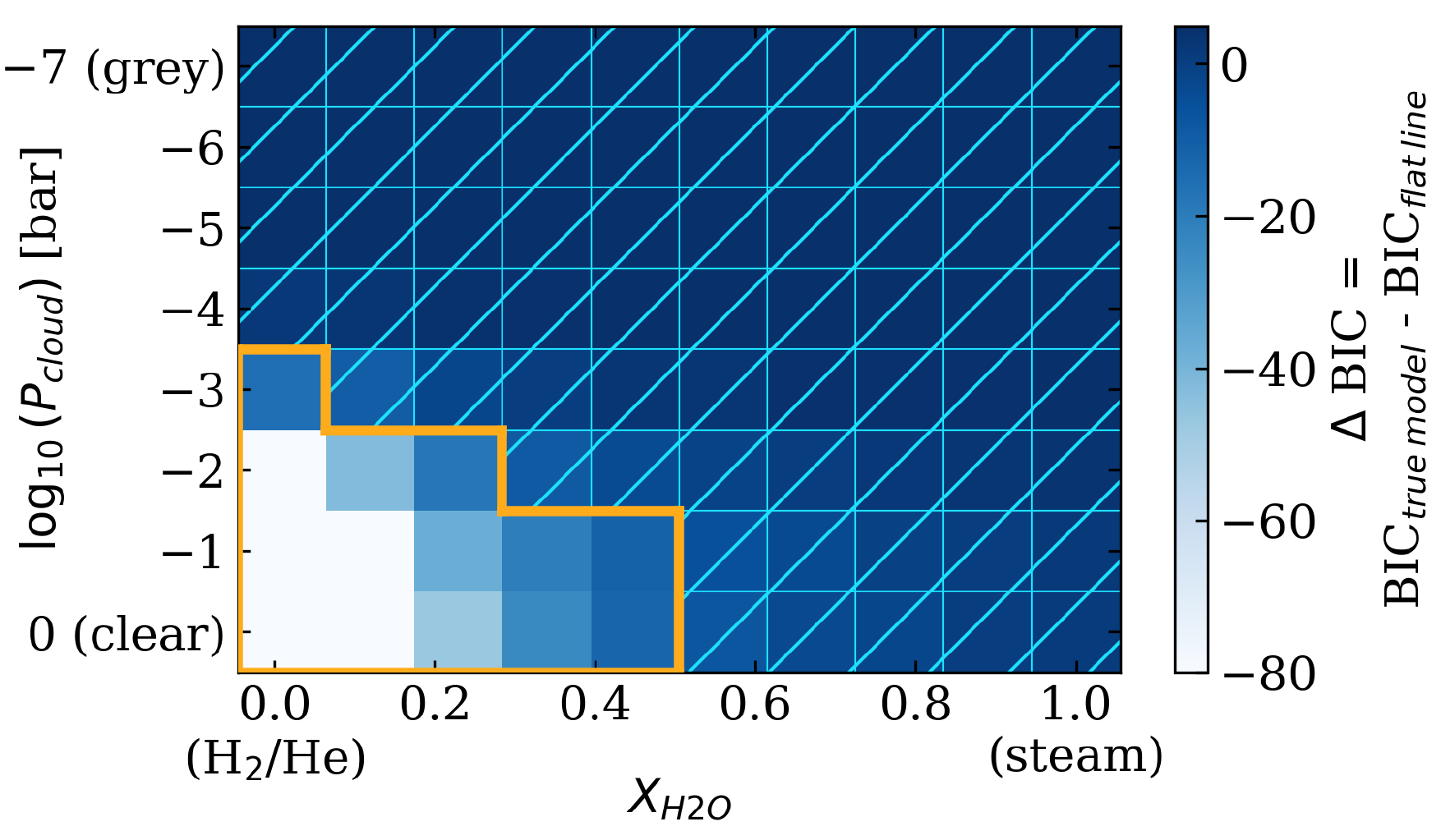}
\caption{{\it Left}: A simulated Ariel Tier 2 transmission spectrum from two transit observations of a GJ~1214~b-size planet, but with a clear, H$_2$/He-dominated atmosphere. {\it Right}: the average detectability map of our target sample as a function of fractional water abundance $X_{\mathrm{H2O}}$ and cloud-deck pressure $P_{\mathrm{cloud}}$. The orange region ($\Delta \mathrm{BIC} < -20$) highlights where we expect to distinguish spectral signatures in transmission from a flat line with two transits.}
\label{fig:bulk_compositions}
\end{figure*}

\begin{figure}
\centering
\includegraphics[width=0.5\textwidth]{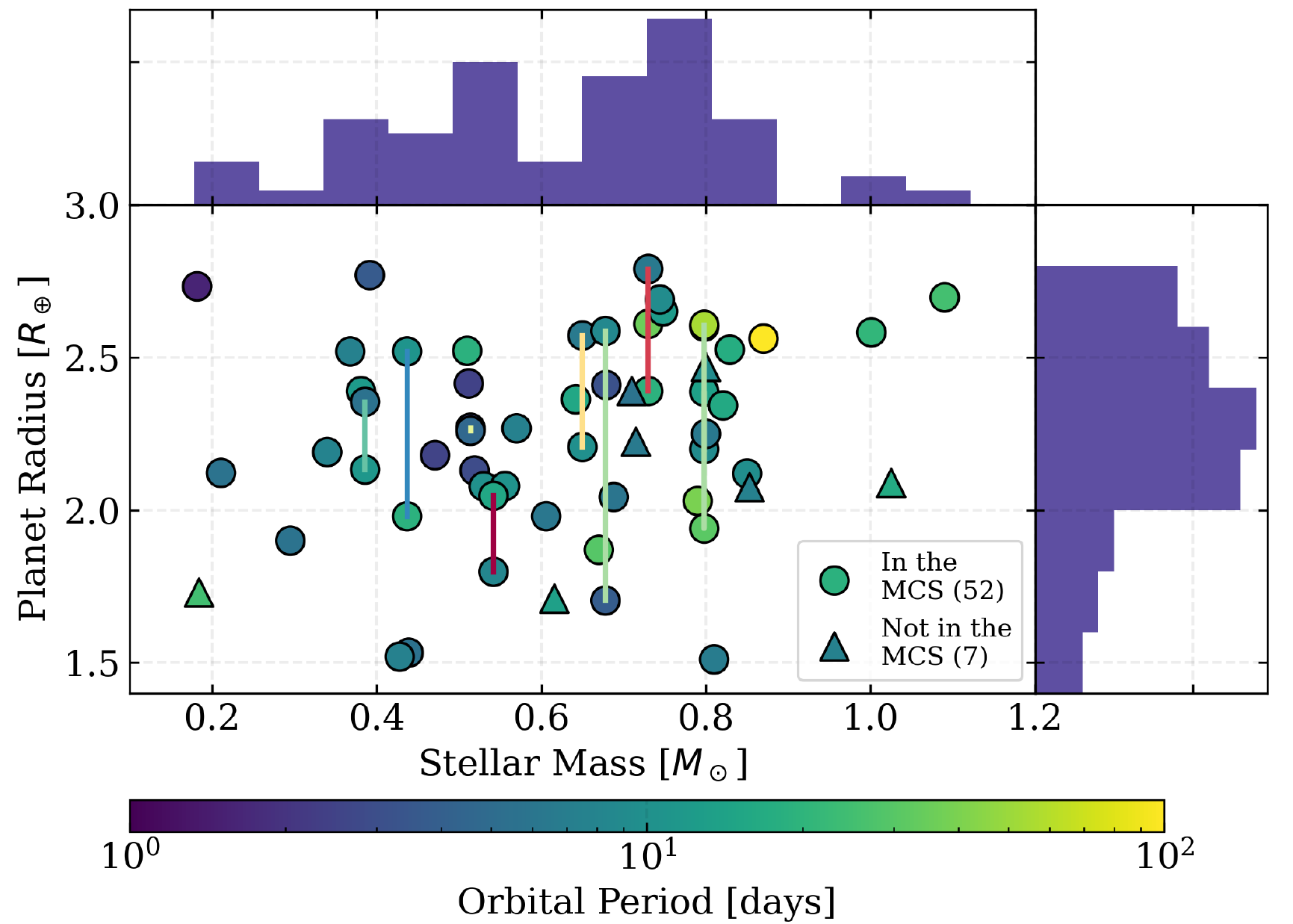}
\caption{The planetary radii, orbital periods, and host stellar masses of our proposed sub-Neptune target list. The spectral types of our full sample span from mid-M to mid-G dwarfs, as is necessary for our experimental design. Members of multi-planet systems are connected by vertical lines and planets in and not in the Mission Candidate Sample (MCS) are depicted by the circle and triangle markers, respectively.}
\label{fig:targets}
\end{figure}

The success of our program rests on our ability to distinguish high mean molecular weight (MMW) atmospheres from flat transmission spectra that are contaminated by clouds. 
We simulate Tier 2 Ariel transmission spectra from two transit observations using a bespoke and host star-dependent adaptation of the Ariel noise model presented in \citet{Mugnai2020} (Fig.~\ref{fig:bulk_compositions}, left panel). We note that we adopt two transits per target based on lessons learned from JWST spectroscopy of small planets, which can produce spurious results and inaccurate inferences on atmospheric compositions \citep{May_2023}. For each planet, we compute the detectability map of spectral features as a function of fractional water abundance $X_{\mathrm{H2O}}$ and cloud-deck pressure by comparing forward models of atmospheric transmission \citep{MacDonald_2023} with a null model (i.e. a flat line). The average detectability map for our final sample is shown in the right panel of Fig.~\ref{fig:bulk_compositions}. We restrict our target list to planets for which we expect sufficient sensitivity with Ariel to detect spectral signatures, at least when assuming a clear atmosphere with low MMW. Our final target list contains 59 sub-Neptunes, of which 50 are currently in the Mission Candidate Sample. The remaining are not included because, although their host stars can be observed, too many visits would be required to achieve Tier 2 data quality. These targets (Fig.~\ref{fig:targets}) orbit a range of host spectral types with masses $M_\star \in [0.18,1.10], M_\odot$ (median $M_\star = 0.65, M_\odot$) and span a range of equilibrium temperatures $T_{\mathrm{eq}} \in [225,870]$ K (median $T_{\mathrm{eq}} = 600$ K). We emphasize that the detectability estimates presented here assume approximately solar atmospheric composition and cloud-free conditions. For planets with higher metallicities or strong cloud opacity (e.g. GJ~1214~b; \citealt{Kempton2023}), the number of transits required to reach Tier~2 precision may be significantly higher. Precisely, our targets are, including those not in the MCS in parenthesis: GJ~1214~b, GJ~3090~b, (HD~219134~c), HD~260655~c, TOI-1064~c, GJ~9827~d, TOI-178~d, HD~110067~c, TOI-2134~b, TOI-2443~b, HD~207496~b, AU~Mic~c, HD~110067~f, TOI-270~c, HD~73583~b, TOI-4438~b, (HD~136352~d), HD~110067~b, TOI-1266~b, HD~97658~b, TOI-2128~b, TOI-836~c, TOI-270~d, TOI-1453~c, G~9-40~b, HD~235088~b, TOI-1201~b, (HD~110067~g), (HD~110067~e), LTT~3780~c, (Kepler-37~d), TOI-836~b, HD~15337~c, TOI-663~b, HD~63433~c, TOI-663~c, TOI-1266~c, TOI-2018~b, TOI-244~b, LHS~1140~b, TOI-2136~b, TOI-1184~b, TOI-260~b, TOI-269~b, TOI-2120~b, TOI-178~e, HD~23472~c, (HD~73583~c), TOI-776~b, TOI-1801~b, TOI-1470~b, (GJ~143~b), Wolf~503~b, K2-3~b, (HD~207897~b), (HD~21520~b), HIP~116454~b, TOI-1260~b, and TOI-776~c. 
 
A robust detection of spectral features is critical to the success of our program such that future target selection activities may attempt to avoid regions of the planetary parameter space where cloud/haze formation is expected to be efficient \citep[e.g.][]{Brande_2024}. Still the non-detections results from cloudy atmospheres will also help us characterize the sub-Neptune population properties. From the current target list, we expect to be able to robustly distinguish between cloudy and high MMW `steam' atmospheres \citep[i.e. $X_\mathrm{H2O}\sim 30$\%;][]{Piaulet_2024} for approximately 37 targets.

\subsection{Disentangling clouds, hazes, and metallicity to reveal atmospheric properties in sub-Neptunes}
\subsubsection{Context}

Atmospheric aerosols, including condensate clouds and photochemical hazes, profoundly shape the transmission spectra of sub-Neptune exoplanets. Observations with JWST have revealed a striking diversity of spectral behavior among planets with otherwise similar physical parameters: some mini-Neptunes show muted or flat spectra indicative of high-altitude opacity \citep[e.g.,][]{Gao2023,Kempton2023,Schlawin2024}, while others exhibit clear molecular features with limited aerosol effects \citep[e.g.,][]{Madhusudhan2023,Jaziri2025b}. A broader, apparently continuous range of spectral morphologies among warm sub-Neptunes \citep[e.g.,][]{Benneke2024,Roy2025,Beatty2024} suggests that aerosol prevalence is not a simple function of planetary size or equilibrium temperature, and that a clearer understanding of aerosol processes is needed to interpret population-level trends.

A major challenge in interpreting transmission spectra is the well-known degeneracy between high atmospheric mean molecular weight (MMW), aerosol opacity, and metallicity. In the infrared alone, muted molecular features can arise either from high MMW atmospheres (e.g., high metallicity steam atmospheres) or from cloud/haze opacity that obscures spectral features \citep[e.g.,][]{Crossfield2017}. This cloud/haze-metallicity degeneracy hampers efforts to connect observed spectra to underlying atmospheric composition and formation history.

Crucially, Ariel's spectral coverage extends into the visible, providing measurements of the short-wavelength scattering slope, which is sensitive to aerosol particle size and composition but largely inaccessible to JWST. The visible data points enable direct constraints on the shape of the cloud/haze opacity and the Rayleigh scattering slope, offering a powerful discriminator between high MMW atmospheres and aerosol opacity. Detecting a steep slope in the visible can signify small-particles hazes, while a flat or muted visible spectrum in combination with infrared features can point toward high metallicity atmospheres with limited aerosol impact.

Photochemical hazes, generated by ultraviolet (UV) irradiation of atmospheric precursors, are expected to be abundant in sub-Neptunes, especially for planets around low-mass stars with elevated UV flux relative to the optical \citep[e.g.,][]{Lavvas2021,Steinrueck2023}. However, predicting haze production from first principles remains challenging due to uncertainties in photochemical pathways, parent molecule abundances, and vertical mixing. Establishing empirical trends in haze prevalence as a function of irradiation, host star spectral type, and atmospheric composition is therefore essential.

Together, these considerations motivate a targeted observational program with Ariel that exploits its combined visible-infrared spectral baseline to break the cloud/haze-metallicity degeneracy and to build population-level constraints on aerosol and photochemical haze production in sub-Neptunes.

\subsubsection{Strategy for observations}

We propose a structured Ariel observational strategy that leverages both the visible spectral coverage and broader sample of sub-Neptunes already planned for other science cases to disentangle aerosols from composition effects.

Ariel'Tier 2 transmission spectra include measurements in the visible (e.g., with the Fine Guidance System and visible photometer channels) that are crucial for constraining the scattering slope induced by aerosols. For each target, combining visible and infrared data will allow retrievals that independently constrain aerosol properties (particle size distribution, scattering slope) and atmospheric composition (metallicity, MMW). This capability is unique among current and planned facilities and cannot be achieved with JWST alone.

Rather than defining a separate target list solely for aerosols, we anticipate that the majority of planets observed for other Ariel key questions (e.g., bulk compositions, chemical tracers) will also inform aerosol studies. Because aerosol prevalence remains difficult to predict a priori, maximizing synergy across science themes ensures broader parameter space coverage. Planets with existing JWST spectra indicating muted features (e.g., GJ 1214 b analogs), those with diverse stellar hosts (from mid-M to G dwarfs), and planets spanning a range of equilibrium temperatures and irradiation levels should be prioritized for Tier 2 coverage when possible.

Stellar heterogeneity (spots, faculae) and activity can imprint spectral slopes in the visible that mimic or obscure aerosol signatures, especially for active M dwarfs. To mitigate this effect: (1) Observations should be scheduled with contemporaneous monitoring of stellar activity indicators (e.g., high-cadence photometry, spectroscopic activity tracers) to identify spot crossing events or rotational modulation. (2) Retrieval frameworks should incorporate models of unocculted starspots and faculae (e.g., \citealt{Rackham2018, Zhang2018, MacDonald2023}) to separate stellar and planetary contributions to the visible slope. (3) For stars with particularly high activity levels, additional transits may be necessary to average over stellar variability and ensure that the measured scattering slope reflects planetary, not stellar, properties.

Recent advances in correcting for stellar contamination, such as the use of multi-band monitoring to constrain spot coverage and temperature contrasts \citep[e.g.,][]{Wakeford2019, Morris2018}, should be integrated into Ariel data analysis pipelines to improve the robustness of aerosol inferences.

Recognizing that aerosol signatures vary across planets and may correlate with irradiation, stellar type, or atmospheric scale height, we recommend a population statistical analysis leveraging the full Ariel sample. This includes planets observed for other science goals whose data quality supports combined visible-infrared retrievals. By comparing trends in scattering slopes, cloud/haze opacities, and retrieved metallicities across this sample, we might be able to quantify how aerosol prevalence depends on planetary and stellar parameters, and identify regimes where photochemical haze production is most efficient.

Ultimately, this strategy has the potential to enable Ariel to break the cloud/haze--metallicity degeneracy and provide the first population-level empirical constraints on photochemical haze production in sub-Neptunes, advancing our understanding of atmospheric processes and informing models of exoplanet atmospheric evolution.

\subsection{What are the magma-envelope signatures of hot sub-Neptunes?}

\subsubsection{Context}

Sub-Neptunes are commonly found exoplanets in the galaxy; yet, there remains a significant uncertainty in understanding their interior structure and composition based on mass and radius. Constraints on their mass and radius suggest that hydrogen and helium constitute only about 0.1--10\% of the planet's mass (by weight). JWST is deciphering the atmospheric composition of sub-Neptunes \citep{Madhusudhan2023,Benneke2024,Davenport2025}, but there is no clear trends in the metallicity or C/O ratio. The highly competitive nature of JWST hampers a sample-level study of sub-Neptunes. The lack of such a survey also poses a challenge for testing theories of planet formation and evolution. The only way forward is to characterize the atmospheres of a significant sample size of sub-Neptunes.

Hot sub-Neptunes with planet equilibrium temperatures exceeding 500~K provide an unprecedented opportunity to probe their atmospheres with Ariel. Recent studies on hot sub-Neptunes and Neptunes provide evidence of haze-free atmospheres that are well suited for characterization efforts \citep{Davenport2025,barat2025}. The hot atmospheres with larger atmospheric scale heights make these planets suited for Ariel observations. Magma oceans on sub-Neptunes have been theorized to persist for more than 1 Gyr \citep{Vazan2018,Tang2025}, and magma-envelope interactions are expected to strongly influence atmospheric composition \citep{Schlichting2022}. Such coupling may leave observable signatures, including SiO at $\sim$5~$\mu$m \citep{Zilinskas2023} and SiH$_4$ at $\sim$4.5~$\mu$m \citep{Misener2023,Charnoz2023,Ito2025}. In addition, high Si/C ratios could signal low metallicity and ongoing magma-envelope exchange \citep{Hakim2026}, while elevated C/N ratios may reflect the selective dissolution of nitrogen-bearing species in magma \citep{Shorttle2024}.
Statistical constraints on metallicity and C/O ratio will help resolve fundamental questions on the formation and evolution of hot sub-Neptunes and Neptunes.

\subsubsection{Strategy for observations}

\begin{figure}
\centering
\includegraphics[width=0.5\textwidth]{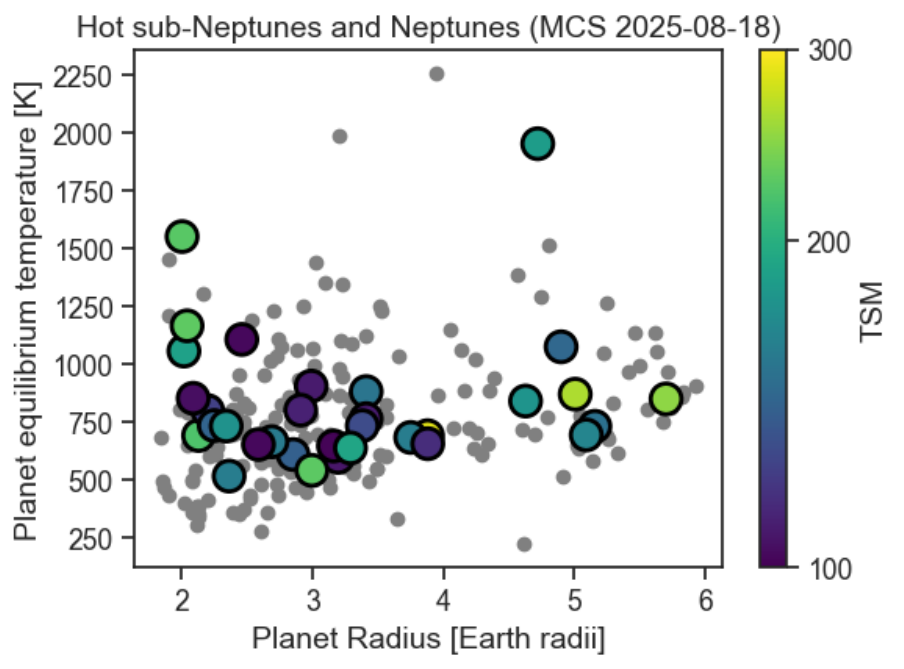}
\caption{Out of 200 hot sub-Neptunes and Neptunes present in the MCS \citep{Edwards2022}, 30 targets will be prioritised based on planet equilibrium temperature ($>$500~K), TSM ($>$100) and J magnitude ($<$9).}
\label{fig:subNeptunesMagma} 
\end{figure}

To address the question of the composition and structure of sub-Neptunes, transmission observations would be the most effective approach. There are about 200 hot sub-Neptunes and Neptunes in the Mission Candidate Sample with a radius of 1.8--6 $R_{\oplus}$ and a host star type of FGKM (Fig.~\ref{fig:subNeptunesMagma}). We prioritize 30 targets based on equilibrium temperatures exceeding 500 K ($>10\times$ Earth's insolation) to select hotter targets, and TSM $>$ 100 and J magnitude $<$ 9, to ensure high-precision ground-based follow-up opportunities.

\subsection{Lava Worlds: existence and composition of their atmospheres} 

\subsubsection{Context}
Lava worlds are among the most extreme rocky exoplanets known: rocky, close-in planets subjected to intense stellar irradiation, with equilibrium temperatures often exceeding 2000 K \citep{Leger2011, Lichtenberg2025}. These planets are typically tidally locked, presenting a permanent dayside to their host stars and a perpetual nightside hidden in cold and darkness. On the dayside, temperatures driven by stellar irradiation are sufficient to maintain global magma oceans \citep{Kite2016}, potentially leading to vigorous outgassing of rocky material and the formation of silicate-rich, metal-bearing atmospheres \citep{vanBuchem2023}. These atmospheres, if present, would be dominated by refractory species such as SiO, Na, Mg, or Fe-bearing vapors, unlike anything seen in the Solar System \citep{Schaefer2009, Miguel2011}.

Despite this theoretical expectation, the atmospheric properties of lava worlds remain largely unknown. Some may retain tenuous atmospheres composed of silicate vapor, while others could possess more substantial volatile-rich atmospheres, possibly enriched by magma ocean outgassing \citep{Zilinskas2021, Zilinskas2022, Zilinskas2023}. It is also unclear what happens on the cold nightside, whether material condenses and is sequestered into the interior, or whether horizontal transport maintains a global atmosphere. It is even plausible that some of these planets have no atmosphere at all, their volatile components long lost to thermal escape and stellar activity.

Recent JWST observations have confirmed the presence of atmospheres in a few lava worlds \citep{Hu2024} yet the majority of data remain unpublished or inconclusive. This emerging picture reinforces the need for systematic, population-level observations to determine whether atmospheres on these planets are common, and if so, what they are made of. While these planets are typically too small and too hot for traditional transmission or emission spectroscopy with Ariel, their extreme thermal environments present a unique opportunity for atmospheric detection via phase curve observations.

\subsubsection{Strategy for observations}
While transmission and emission spectroscopy of lava worlds is beyond Ariel's sensitivity limits due to their small sizes, thermal phase curve observations offer a promising path forward. Ariel's ability to conduct broadband photometry and low-resolution spectroscopy across multiple epochs makes it well-suited to probe the longitudinal temperature structure of these tidally locked planets. From this, one can infer key properties such as the dayside and nightside temperatures, heat redistribution efficiency, and even optical phase curves for albedo estimates.

By measuring thermal phase curves, Ariel can test whether these planets possess atmospheres at all. A large temperature contrast between day and night suggests a bare rock with no significant atmosphere to redistribute heat. Conversely, a muted contrast could indicate the presence of an atmosphere capable of transporting energy from the dayside to the nightside. In brighter systems or those with multiple phase curves, Ariel may even be able to begin constraining atmospheric composition, particularly in cases where silicate-bearing species or thermal inversions alter the emission spectrum.

To maximize the scientific return, Ariel should prioritize short-period rocky planets with high equilibrium temperatures and bright host stars (e.g. J $<$ 9). A small sample of well-chosen targets, ideally overlapping with those observed by JWST (e.g., TOI-1807~b, TOI-2260~b, TOI-431~b, TOI-2431~b, TOI-6255~b; Proposal 8864, PI: L. Dang), would offer not only wavelength complementarity but also enable joint constraints on temperature structure and atmospheric composition. Furthermore, Ariel phase curves will be significantly more efficient than previous Spitzer observations, with preliminary estimates suggesting that 1-2 Ariel phase curves are equivalent to $\sim$10 from Spitzer in signal-to-noise.

Ultimately, a population study of lava worlds with Ariel's phase curve capabilities will address a fundamental question: do these planets retain atmospheres, and if so, what are they composed of? By answering this, Ariel will be able to shed light on atmospheric survival, surface-atmosphere interactions, and the boundary between rocky exoplanets with and without volatile envelopes in the most extreme environments known.

\subsection{Observing Temperate Exoplanets with Ariel}\label{sec:temperate}

\subsubsection{Context}

While Ariel is primarily designed to study warm and hot exoplanets (with equilibrium temperatures above 500 K), recent advances in both ground- and space-based instrumentation have enabled the detection of a growing number of temperate exoplanets. These include gas giants and sub-Neptune-size planets orbiting F, G, and K stars, with orbital periods between 100-300 days and temperatures ranging from 250 to 500 K, particularly highlighted by TESS discoveries \citep{encrenaz2018}.
Temperate exoplanets, especially those with radii less than 4 Earth radii, fall into two main categories: sub-Neptunes and super-Earths. Differentiating between these types is critical for understanding planetary formation and evolution. However, this distinction cannot be made based on bulk density alone, it requires atmospheric characterization via transit spectroscopy. By probing their atmospheric composition, we not only broaden our understanding of exoplanet diversity but also bridge the gap between known exoplanets and those in our own solar system, paving the way toward studying potentially habitable worlds.
One advantage of targeting temperate planets is their atmospheric clarity: thermochemical models predict fewer condensates in the [300-500] K range, simplifying spectral interpretation compared to hotter planets. Though Ariel's mission goals focus on warmer targets, simulations show that it is feasible to detect and study several temperate exoplanets, particularly larger ones (5-15 Earth masses) with hydrogen-rich atmospheres. Using the ArielRad code \citep{Mugnai2020} and Mission Candidate Sample \citep{Edwards2022}, about 15 temperate planets have been identified as observable in Ariel's Tier 2 spectroscopic mode, including gas giants, large Neptunes, and intermediate-mass planets \citep{encrenaz2022}.

\begin{figure}
\centering 
\includegraphics[width=0.5\textwidth]{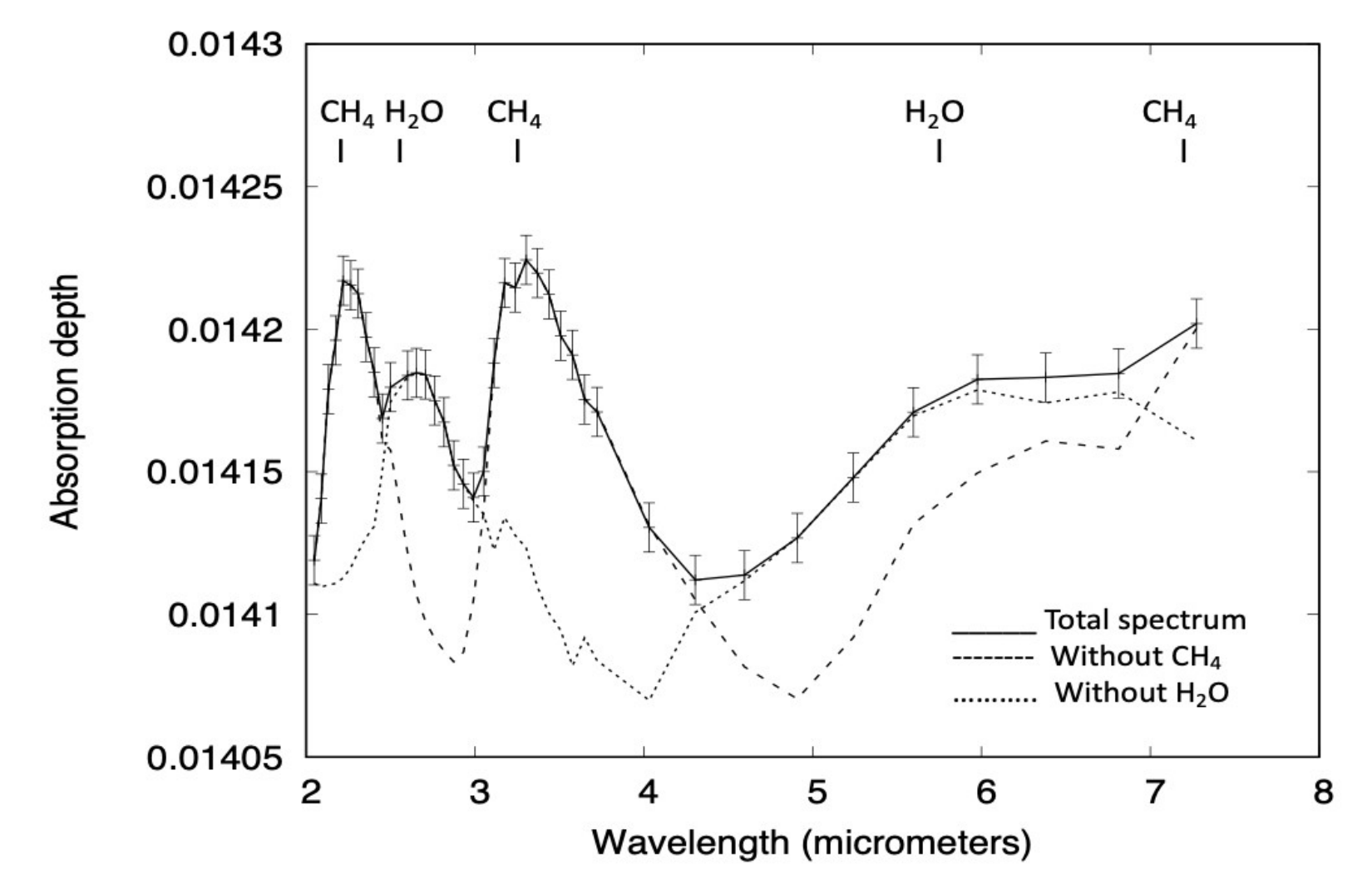}
\caption{Solid black line: A typical spectrum of a temperate Jupiter (T$_\mathrm{eq}$~=~400~K) as seen in transmission in front of a K0-type star, with a spectral resolution corresponding to Tier 2 (R~= ~50 below 4 $\mu$m and 15 above, Ariel Redbook, Table 3-3). Dashed line: the same spectrum without the CH$_4$ contribution; dotted line: the same spectrum without the H$_2$O contribution. Calculations are performed with the Planetary Spectrum Generator \citep{villanueva2018}. The error bars are calculated with the ArielRad code \citep{Mugnai2020} for 16 transits (assuming a planet located at 50 pc), and correspond to a mean S/N of 7 over the spectrum. Figure from \protect\cite{encrenaz2022}.}\label{fig:temp_exo}
\end{figure} 

\subsubsection{Strategy for observations}
\begin{figure*}
\centering
\includegraphics[width=0.49\textwidth]{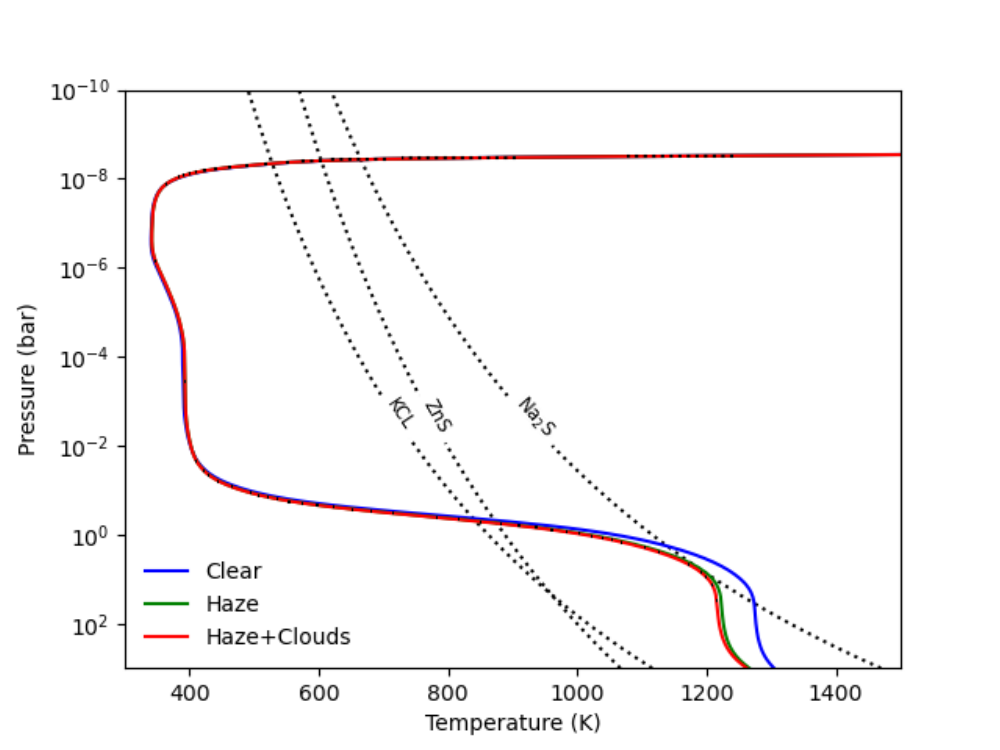}
\includegraphics[width=0.49\textwidth]{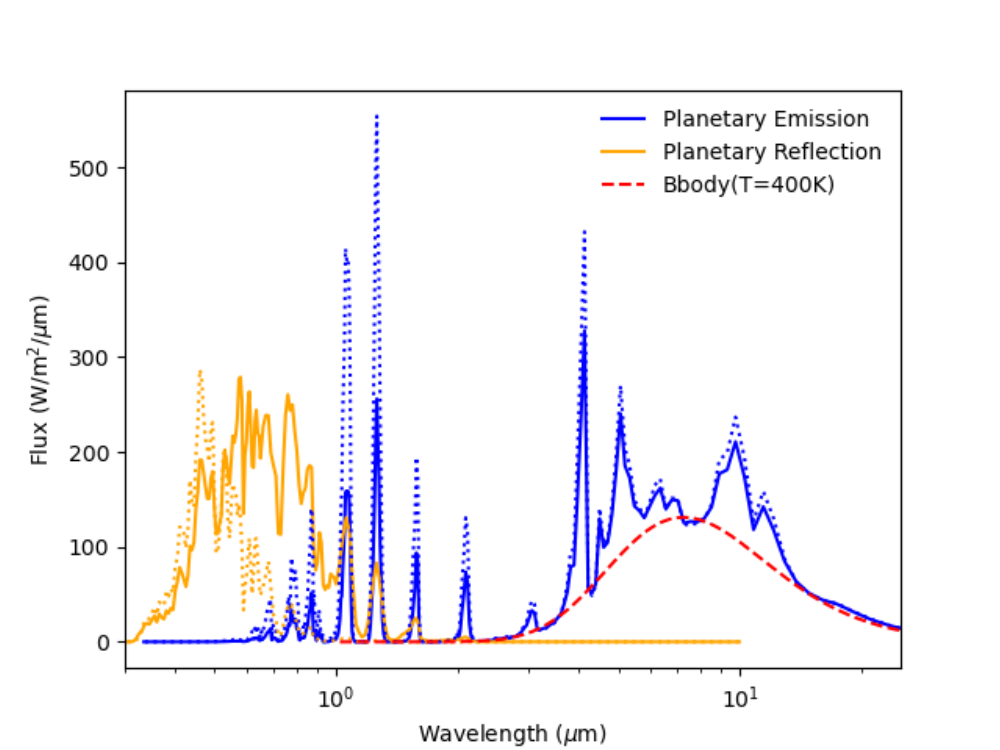}
\caption{Left: Impact of haze and clouds on the atmospheric temperature profile. Right: Outgoing radiation from the atmosphere of TOI-1759 b, with reflected and emitted contributions highlighted. The dotted lines are the corresponding contribution for the clear atmosphere case and the dashed line presents the thermal emission from a 400 K black body.}\label{fig:temp_exo2} 
\end{figure*}

Hydrogen-rich exoplanets, from Jupiters to sub-Neptunes, are especially promising for transit spectroscopy because their extended atmospheres produce stronger transmission signals, roughly ten times greater than those of heavier-molecule-dominated super-Earths. Key spectral features for atmospheric characterization fall within the visible to near-infrared range, with strong absorption bands from CO (4.7 $\mu$m), CO$_2$ (4.3 $\mu$m), CH$_4$ (3.3 $\mu$m), and H$_2$O (2.7 $\mu$m), along with notable features from NH$_3$ and HCN near 3 $\mu$m.
Infrared transmission spectra were computed across different stellar types, showing that temperate Jupiters and sub-Neptunes with short-to-intermediate orbital periods (tens of days) can be observed by Ariel at distances up to 50 pc (for Jupiters) and 25 pc (for sub-Neptunes), assuming Tier 2 mode (R $\simeq$ 50 below 4 $\mu$m and R $\simeq$ 15 above 4 $\mu$m)(Fig.~\ref{fig:temp_exo}). In contrast, temperate super-Earths remain beyond Ariel's Tier 2 capabilities.
Target selection depends on both planetary (mass, radius, equilibrium temperature) and stellar (radius, brightness) characteristics. Two key metrics guide this process \citep{encrenaz2023}:
\begin{itemize}
    \item Transmission Spectroscopic Metric (TSM): a measure that increases with planet size and temperature, and decreases with mass, stellar radius, and brightness.
    \item T2 Factor: the number of transits needed for spectral detection with Ariel (lower is better), also applicable to JWST and ground-based telescopes.
\end{itemize}
Ideal targets exhibit high TSM and low T2 values, typically, TSM $\ge$ 50 and T2 $\le$ 20. For example, the sub-Neptune TOI-1759~b has been identified as a viable Tier 2 target for Ariel \citep{lavvas2024}, and atmospheric models have been developed for it (Fig.~\ref{fig:temp_exo2}). 
An interesting characteristic of these cooler environments is the increasing contribution of sulfuric chemistry in their atmospheres with the possible detection of OCS, CS$_2$, and other organosulfur photochemical products, as well as the possible inclusion of this element in the photochemical hazes of such atmospheres. Observations of such planets would therefore complement those proposed in Sect.~\ref{sec:sulfur}, helping to build a more complete picture of the chemical diversity in exoplanet atmospheres.

\begin{table*}
\caption{Estimated parameters of the primary transit signal of a small Neptune-like exoplanet transiting around a star of spectral type between G2 and M8, with an albedo a = 0.3, assuming either a fast rotator (columns 5 and 7) or a tidally locked object (columns 6 and 8). $R_\ast$, $M_\ast$ and $T_\ast$ are the radius, the mass and the effective temperature of the star. Best values are in bold, following \protect\cite{encrenaz2023}.}     
\label{tab:sample}      
\centering          
\begin{tabular}{l r r r r r r r r c r r r}
\hline\hline  
Spectral & $R_\ast$  & $M_\ast$& $T_\ast$  & D$_\mathrm{fast}$ & D$_\mathrm{slow}$   & P$_\mathrm{fast}$  & P$_\mathrm{slow}$ & $\tau$ & A & Nb  & Total time \\
type & ($R_{\odot}$) & ($M_{\odot}$)& (K) &  (AU) & (AU)  &  (d)  &  (AU) & (h) &  & transits &  (y) \\
(1) & (2) & (3) & (4) & (5) & (6) & (7) & (8) &  (9) & (10) & (11) & (12) \\
\hline                    
   G2  &  1.0 & 1.0 &  5770 & \textbf{0.41}  & 0.57 & \textbf{95} & 158 & 8.32 & 6.31(-5) & 8 (12.5 pc) & 2.08 (12.5 pc)\\
     &   &  &   &   &  &  &  &  &  & 11 (25 pc) &  2.86 (25 pc)\\
 
   G5   &  0.93 & 0.93 & 5641 & \textbf{0.36} & 0.5 & \textbf{81} & 136 & 7.51 & 7.30(-5) & 6 (12.5 pc)  & 1.33 (12.5 pc) \\
    &   &  &   &   &  &  &  &  &  &  10 (25 pc) &  2.22 (25 pc)\\
 
   K0   &  0.85 & 0.78 & 4977 &  \textbf{0.26} & \textbf{0.36} & \textbf{53} & \textbf{89} & 6.22 & 8.73(-5) & 5 (12.5 pc)  & 0.73 (12.5 pc) \\
      &   &  &  &   &  &  &  &  &  &  8 (25 pc) &  1.16 (25 pc)\\
   K5   &  0.74 & 0.59 &  4242 &  0.16 & \textbf{0.23} & 29 & \textbf{48} & 5.54 & 1.15(-4) & 4 (12.5 pc) & 0.53 (12.5 pc)\\
    &   &  &   &   &  &  &  &  &  & 8 (25 pc)&  1.05 (25 pc)\\

   M0   &  0.63 & 0.47 & 3642 &  0.10 &  \textbf{0.14} & 17 & \textbf{28} & 4.52 & 1.59 (-4) & 3 (12.5 pc)  & 0.23 (12.5 pc) \\
   &   &  &   &   &  &  &  &  &  &  8 (25 pc) &  0.61 (25 pc)\\

   M5    &  0.32 & 0.21 & 3041 & 0.02 &  \textbf{0.03} & 2.8 & \textbf{4.5} & 1.72 & 6.16(-4) & 2 (12.5 pc)  & 0.02 (12.5 pc) \\
  &   &  &   &   &  &  &  &  &  &  7 (25 pc) & 0.09 (25 pc)\\

   M8   & 0.13 & 0.10 & 2691 & 0.01 & \textbf{0.02} & 1.3 & \textbf{2.3} & 0.54 & 3.73 (-3) & 2 (12.5 pc) & 0.01 (12.5 pc) \\
 &   &  &   &   &  &  &  &  &  &  9 (25 pc) & 0.06 (25 pc)\\

\hline                  
\end{tabular}
\end{table*}

\section{Conclusions}

This white paper has outlined the key science cases motivating the Chemistry Working Group of the Ariel mission. By exploring the detectability and diagnostic potential of a broad range of chemical species, including sulfur- and phosphorus-bearing molecules, photochemical by-products, and species sensitive to metallicity, C/O ratio, and stellar activity, we have provided a science-driven framework to support Ariel's target selection.

For each science case, we have proposed observational strategies and identified specific targets that maximize Ariel's capability to address current gaps in our understanding of exoplanet atmospheric chemistry. In particular, we emphasized the importance of population-level studies to reveal chemical trends, and the value of targeted observations to test theoretical predictions.

Ariel's large and diverse planet sample, coupled with its broad spectral coverage, offers a unique opportunity to investigate the role of disequilibrium processes, elemental reservoirs, and photochemical responses to external forcing (e.g. stellar flares). The recommendations outlined in this report aim to ensure that the Chemistry WG's priorities are well represented in the final Mission Candidate Sample, so that Ariel can deliver transformative insights into the chemistry, diversity, evolution, and origins of planetary atmospheres.

\section*{Aknowledgements}
O.V. acknowledges funding from the ANR project `EXACT' (ANR-21-CE49-0008-01) and from the Centre National d'Etudes Spatiales (CNES).
Y.M. acknowledges support from the European Research Council (ERC) under the European Union's Horizon 2020 research and innovation programme (grant agreement no. 101088557, N-GINE).
A.Y.J. has received funding from the European Research Council (ERC) under the ERC OxyPlanets projects (grant agreement No.101053033). 
D.R.L. would like to acknowledge that this publication has emanated from research conducted with the financial support of Taighde {\'E}ireann - Research Ireland under Grant number 21/PATH-S/9339.
K.H. acknowledges the FED-tWIN research program STELLA (Prf-2021-022) funded by the Belgian Science Policy Office (BELSPO) and the research grant G014425N funded by the Research Foundation Flanders (FWO).
Z.M., through Centro de Qu\'imica Estrutural, acknowledges the financial support of Funda{\c c}{\~a}o para a Ci{\^e}ncia e Tecnologia (FCT) to projects UIDB/00100 and UIDP/00100, and through Institute of Molecular Sciences, to project LA/P/0056. 
E.H. and R.V. were supported by a Science and Technology Facilities Council Small Award [ST/Y00261X/1].
M.Z. was supported by a UKRI Future Leaders Fellowship MR/T040866/1.
H.R.H. acknowledges support from PEPR Origins (`X MARKS the SPOT').

\section*{Data Availability}

All data used in this article were generated within the Chemistry Working Group of the Ariel mission and are available from the corresponding author upon request.
 

\section*{Conflict Of Interest}

Authors declare no conflict of interest.



\bibliographystyle{rasti}
\bibliography{biblio}






\bsp	
\label{lastpage}
\end{document}